\documentclass[superscriptaddress,twocolumn,showkeys,aps,prb,reprint]{revtex4-1}
 
\usepackage{graphicx}
\usepackage{dcolumn}
\usepackage{bm}
\usepackage{float}
\usepackage{color}
\usepackage{amsmath}
\usepackage{calligra}
\usepackage[T1]{fontenc}
\usepackage{tablefootnote}

\usepackage{soul,xcolor}
\usepackage{latexsym}
\usepackage{bbm}
\usepackage{amsfonts}
\usepackage{amssymb}
\usepackage{hyperref}
\hypersetup{colorlinks=true, citecolor=blue}
\setstcolor{red}

\usepackage{physics}
\usepackage{booktabs}

\usepackage{soul,xcolor}
\usepackage{calrsfs}
\usepackage{mathrsfs}
\usepackage{ulem}
\usepackage{tabularx}
\usepackage{siunitx}
\usepackage{array}

\begin{document}

\title{Spin-orbit coupling effects in single-layer phosphorene}

\author{Mayra Peralta}
\email{Mayra.Peralta@cpfs.mpg.de}
\affiliation{Max Planck Institute for Chemical Physics of Solids, 01187 Dresden, Germany}
\affiliation{Yachay Tech University, School of Physical Sciences and Nanotechnology, 100119, Urcuqui, Ecuador}

\author{Dennis A. Freire}
\affiliation{Yachay Tech University, School of Physical Sciences and Nanotechnology, 100119, Urcuqui, Ecuador}%

\author{Rafael Gonz\'alez-Hern\'andez}
\affiliation{Departamento de F\'isica y Geociencias, Universidad del Norte, Km.\,5 V\'ia Antigua Puerto Colombia, Barranquilla 080020, Colombia}

\author{Francisco Mireles
}
\email{fmireles@ens.cnyn.unam.mx}
\affiliation{Departamento de F\'isica, Centro de Nanociencias y Nanotecnolog\'ia, Universidad Nacional Aut\'onoma de M\'exico, Apdo. Postal 14, C.P. 22800, Ensenada B.C., M\'exico}
 
\date{\today}

\begin{abstract}
The electronic band structure of monolayer phosphorene is thoroughly studied by considering the presence of spin-orbit interaction. We employ a multiorbital Slater-Koster tight-binding approach to derive effective $\bm{k}\cdot \bm{p}$-type Hamiltonians that describes the dominant spin-orbit coupling (SOC) effects of the Rashba and intrinsic origin at the high $\Gamma$ and $\rm{S}$ high symmetry points in phosphorene. In the absence of SOC effects  a minimal admixture of $p_z$ and $p_y$ atomic orbitals suffices to reproduce the well known anisotropy of highest valence and the lowest conduction bands at the $\Gamma$-point, consistent with density functional theory (DFT) and $\bm{k}\cdot \bm{p}$ methods. In contrast, the inclusion of the $p_x$ and $s$ atomic orbitals are rather crucial for an adequate description of the SOC effects in phosphorene at low energies, particularly at the $\rm{S}$-point. We introduce useful analytical expressions for the Rashba and intrinsic SOC parameters in terms of the relevant Slater-Koster integrals. In addition, we report simple formulas for the interband dipole-strenghts,  revealing the nature of the strong anisotropic behavior of its lower bands. Our findings can be useful for further studies of electronic and spin transport properties in monolayer phosphorene and its nanoribbons. 
\end{abstract}

\maketitle
 
\section{Introduction}

Single-layer phosphorene has emerged as an appealing and  promissing two-dimensional semiconductor. This is due to its unique electronic, optic, thermo-electric, and mechanical properties, all exhibiting a highly anisotropic character\cite{Takao1981, Koenig2014, Li2014,Castellanos2014, Rodin2014,Sodagar2022}. Among its prominent optoelectronic features are its direct energy band gap at the $\Gamma$-point close to $2.1$\,eV,\cite{Lian2014,Wang2015,Kurpas2018,Junior2019,Frank2019} its very high $p$-type mobility ($\sim\!\! 10^3$\,cm$^2$V$^{-1}$s$^{-1}$), and its highly anisotropic bandstructure near the Fermi energy. The latter contributing to its large anisotropic interband dipole couplings\cite{Zhang2014,Li2019}. In addition, few  layer phosphorene have shown also a layer-sensitive band gap with an extraordinary enhanced photoluminescence intensity\cite{Zhang2014}. All these features, make of phosphorene an excellent candidate for applications in  optoelectronic, photonic, and energy saving two-dimensional devices \cite{Kou2015, Liu2015, Zhang2020, Farghadan2020,Avsar2020,Sierra2021,Zhang2024}.

Recent studies have also targeted phosphorene as a suitable material for spintronics applications owing its relatively weak spin-orbit coupling of its phosphorous atoms \cite{Avsar2017, Kurpas2018, Rahmani2020}.
High spin lifetime has been measured in phosphorene at room temperature  ($\sim\!\!0.7$\,ns)\cite{Avsar2017}, which together with its large spin diffusion lengths leads to coherent spin-polarized transport covering micrometric distances ($\sim\!\! 2.5\,\mu$m).  
Despite the relatively weak spin-orbit effects, phosphorene experiences unavoidable spin scattering events that hinder the control of coherent spin transport. These events stem from two primary sources of spin scattering induced by spin-orbit interactions in phosphorene.
They are of intrinsic and extrinsic nature, leading to the known Elliot-Yafet, and the D’yakonov-Perel’ spin-relaxation mechanisms, respectively. Recent estimates of the spin-relaxation lifetimes yields strongly anisotropic relaxing times in the plane of phosphorene, showing field and carrier dependence for both types of  spin relaxation mechanisms. 
Moreover, intriguing interplays are anticipated to emerge under moderate electric fields, though with dominant (strongly anisotropic) D’yakonov-Perel’ spin-scattering at high electric fields\cite{Kurpas2016}. The origin of such anisotropies was associated to the expected highly anisotropic Rashba splitting, although no explicit expressions for the anisotropic Rashba-SOC Hamiltonians was provided. The spin-dependent electronic properties of phosphorene have been also recently addressed introducing phenomenological anisotropies of the Rashba coupling strength weighted by the ratio of the effective masses along the $x-$ and $y-$ direction but without further physical insights\cite{Popovic2015}. As for the intrinsic type of spin-orbit interaction, an earlier DFT calculation\cite{Kurpas2016} identify that its dominant contribution arises at the ${\rm S}$ high symmetry point within the Brillouin zone, yielding a uniform band spin-splitting of about $17.5$\,meV at the lowest conduction band. Nevertheless, the nature of the atomic hibridizations participating in the spin splitting was not examined.   

Motivated by these studies, our foremost goal is to construct a minimal multi-orbital tight-binding model for two-dimensional phosphorene valid under spin-orbit effects. This model will serve as the framework for deriving effective Hamiltonians that accurately characterize the electronic properties of phosphorene's lowest energy bands. Through this approach, we aim to deepen our understanding of the inherent anisotropies within its energy bands and elucidate the interplay of SOC effects in single-layer phosphorene.
 
We began our study in the absence of SOC using a multiorbital tight-binding model consisting of a natural basis of the admixture of rotated $p_z$ and $p_y$-like atomic orbitals. This allow us to derive an effective Hamiltonian that reproduces the main characteristics of the anisotropy of its electronic band-structure near the Fermi energy as dictated by DFT calculations. Analytical expressions of the effective masses in terms of the Slater and Koster parameters were also derived, offering further physical insights on the origin of the large anisotropy of the conduction and valence band effective masses of phosphorene around the $\Gamma$ high symmetry point within the Brillouin zone. Subsequently, within the same tight-binding footing, a model that incorporates the $s$, and $\{p_x,p_y,p_z\}$ atomic orbitals and the effects of the SOC is introduced. Such model leads to an eight-band effective Hamiltonian that at the  ${\rm \Gamma}-$point describes the predicted anisotropic Rashba coupling in phosphorene. We were able to quantify the ratio of such anisotropy. It is shown that at moderate external electric fields ($\sim3$\,V/nm) and typical carrier densities of $10^{12}$\,$cm^{-2}$ the  strength of the Rashba parameter along the $\Gamma$-Y direction is around 20 times smaller than the one of the $\Gamma$-X direction, and that the the range of the induced spin-splittings is in the order of a few $\mu$eV.  On the other hand, and 
in  contrast with what it occurs near the $\Gamma$-point, the intrinsic type of spin-orbit coupling is the dominant interaction at the $\rm{S}$-point, whiles the Rashba-type is rather small, in agreement with our DFT calculations. We also found that an intrinsic SOC induced spin-splitting is developed at the $\rm S$-point of phosphorene of about $20.4$\,meV. Finally, we present simple formulas for the interband dipole-strenghts,  revealing the origin of the strong anisotropic behavior of its lower bands.

\section{Phosphorene Tight-Binding Model without spin-orbit effects}\label{TBH}

We first analyze the tight-binding model for monolayer phosphorene neglecting the spin-orbit coupling, in a later stage its effects will be  incorporated on equal footing. We start the model by defining the Hamiltonian operator describing the kinetic energy of the electrons and the energy potential $V_{at}$ produced by the  ions of the  2D crystal lattice of phosphorene,
\begin{equation}\label{Hamileq}
\hat{H}_{0}=-\frac{\hbar^2}{2m}\nabla^{2}+\sum_{ij}V_{at}\left ({\bf R}_{i} - {\bf R}_{j} \right), 
\end{equation}
\noindent which within the two-center tight-binding (Slater-Koster)  approximation, only involves the atoms $i/j$  at the vector positions 
${\bf R}_i/{\bf R}_j$, as well as its first atomic neighbors. Considering all the outer shell $3s$ and $3p$ orbitals of each phosphorus atom yields a $16\times16$ Hamiltonian matrix; that is, four atomic orbitals for each of the four distinct   A(A$^{\prime}$) and B(B$^{\prime}$) atomic sites in the unit cell (Fig.1(e,f)). We are interested however, to set up the minimal multi-orbital model which enable us to capture the main features of the lowest energy bands at the highest symmetry points. The proper selection of such atomic orbitals are dictated by the most highly occupied atomic states at the symmetry points of interest. For this we rely on our first principles calculations implemented by DFT (See appendix E for details). Fig.1(c) shows the calculated electronic band dispersion of monolayer phosphorene and Fig.1(a) presents the calculated projected local density of states (LDOS). The projected LDOS shows that the highest valence and minimum conduction bands near the $\Gamma$ point, for instance, have a strong contribution coming from the $p_z$-orbitals, with a moderate contribution  originated by the $p_y$-orbitals,  and a very weak or negligible presence of the $p_x$ and $s$ orbitals (Fig.1(b)). Therefore,  a reasonable starting point to construct a minimal tight-binding model near the $\Gamma$ symmetry point can be chosen with a subspace spanned only by the $p_z$ and $p_y$ atomic orbitals. This is also consistent with group theory symmetry arguments, dictating that in nonsymmorphic group $D_{2h}$ crystals as phosphorene, those states belonging to the $\Gamma_{z-}$ and $\Gamma_{y+}$ character are the only ones that are symmetry allowed at the $\Gamma$-point.\cite{Appelbaum2014}

\begin{figure*}
\hspace{-0.5cm}
\includegraphics[width=17.0cm]{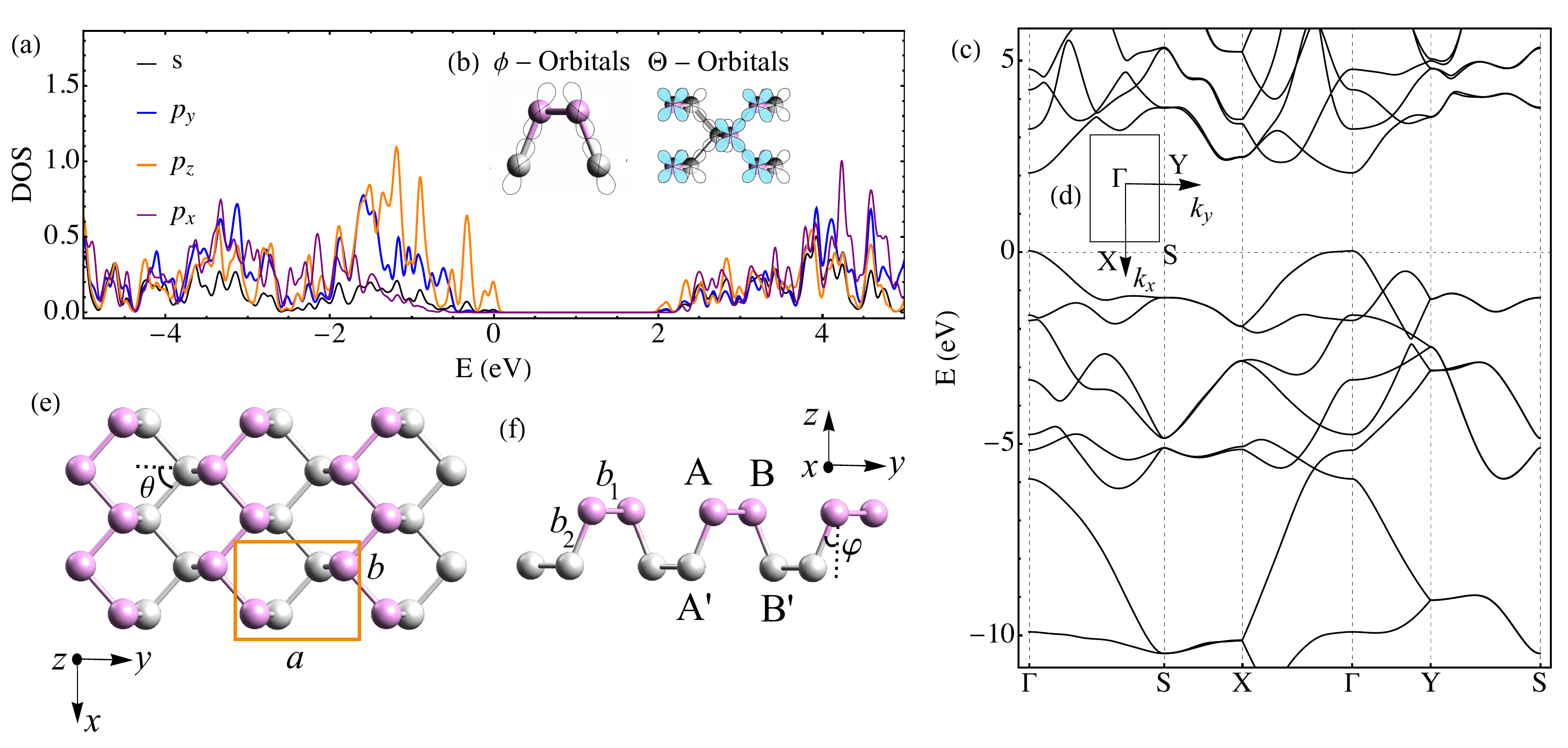}
    \caption{(a) Projected LDOS for the $s$, $p_x$, $p_y$, and $p_z$ orbitals.
    (b) Depiction of the  of $p$-orbitals of phosphorene.  
    (c) Illustration of the band structure of monolayer phosphorene unfolded into the full Brillouin zone obtained by {\it ab initio} DFT calculations. 
    Inset (d), first Brillouin zone of the phosphorene lattice showing its high symmetry points $ \Gamma$, $\rm{X}$, $\rm{Y}$, and $\rm{S}$. (e) Phosphorene's crystalline structure showing the rectangular primitive lattice with dimensions $a$ and $b$, being $\theta$   the angle of the A(A$^\prime$)--B(B$^\prime$) bond with the $y$-axis. (f) Side view, where the four atoms of the basis labeled as A, A$^\prime$, B, and B$^\prime$ are identified. The angle of the  $\hat{z}$-axis with the A(B)--A$^\prime$(B$^\prime$)bond is defined as $\varphi$. The parameters $b_1$ and $b_2$ are the A(A$^\prime$)--B(B$^\prime$) and A(B)--A$^\prime$(B$^\prime$) atomic distances, respectively.
    }
    \label{NoSOC1}
\end{figure*}
Furthermore, given the atomic spatial configuration of the phosphorous atoms in phosphorene (Fig.1(b,f)), we find it useful to introduce a basis set composed of rotated $p_z$ and $p_y$ orbitals defined as,
\begin{equation}\label{basisphi}
\begin{aligned}
|\phi^{\rm A/A'}\rangle  &= -m_{z}|{p_{z}}^{\rm A/A'}\rangle - m_{y}|{p_{y}}^{\rm A/A'}\rangle, \\
|\phi^{\rm B/B'}\rangle  &= -m_{z}|{p_{z}}^{\rm B/B'}\rangle + m_{y}|{p_{y}}^{\rm B/B'}\rangle.
\end{aligned}
\end{equation}
\noindent where $m_y = \sin(\varphi)$ and $m_z = \cos(\varphi)$. Now, since we will assume electron  hoppings occurring only between the first nearest neighbors for each of the four atoms of the unit cell, thus solely the Slater-Koster overlapping integrals between the  $\rm{A-A'}$, $\rm{A-B}$, $\rm{B-B'}$, and $\rm{A'-B'}$ atomic sites  will be needed. Next-nearest neighbors and beyond will be neglected as they are expected to decay exponentially with the interatomic distance.\cite{Menezes2018} It can be shown that the matrix elements that govern the electron hopping energy between $\phi^{\rm{A/B}}$  orbitals from the atomic sites A(B) to the $\phi^{\rm{A'/B'}}$ of the  $\rm{A'(B')}$ atoms form a $\sigma$ type bond, that is, all these matrix elements reduce to $\langle\phi^{\rm{A/B}}|\hat{H}_{0}|\phi^{\rm{A'/B'}}\rangle=V_{pp\sigma}$ with vanishing $\pi$-bonding. Here we have used the conventional Slater-Koster relations 
$\langle p_{\mu}^{\rm{A/B}}|\hat{H}_{0}|p_{\mu}^{\rm{A'/B'}}\rangle$ and $\langle p_{\mu}^{\rm{A/B}}|\hat{H}_{0}|p_{\nu}^{\rm{A'/B'}}\rangle$, as defined in Table\,\ref{SKdefinitions}  with the indices $\{\mu,\nu\}=\{x,y,z\}$. 


\begin{table}
\begin{tabular}{|l|c|}
\hline\hline
Matrix element & in terms of the SK  parameters \\
\hline   
$\langle s^{\rm{A/B}}|\hat{H}_{0}|s^{\rm{A'/B'}}\rangle$ \quad    & $V_{ss\sigma}$\\
$\langle s^{\rm{A/B}}|\hat{H}_{0}|p_{\mu}^{\rm{A'/B'}}\rangle$ \quad    & $n_{\mu}V_{sp\sigma}$\\
 $\langle p_{\mu}^{\rm{A/B}}|\hat{H}_{0}|p_{\nu}^{\rm{A'/B'}}\rangle$ \quad   &   $-n_{{\mu}}n_{{\nu}}(V_{pp\pi}-V_{pp\sigma})$ \\
 $\langle p_{\mu}^{\rm{A/B}}|\hat{H}_{0}|p_{\mu}^{\rm{A'/B'}}\rangle$ \quad   & \quad 
  $n_{{\mu}}^{2}V_{pp\sigma}+(1-n_{{\mu}}^{2})V_{pp\pi}$  \\
\hline \hline
\end{tabular}
\caption{Some Slater-Koster (SK) hopping integrals definitions used in this work, with the indices $\{\mu,\nu\}=\{x,y,z\}$. The cosine directors are given by $n_x=\sin{\theta}\sin{\varphi}$, $n_y=\cos{\theta}\sin{\varphi}$ and $n_z=\cos{\theta}$.}
\label{SKdefinitions}  
 \end{table}

Therefore, the simplest tight-binding Hamiltonian in the basis given by $\{\ket{\bm \phi} \}=\{ \ket{\phi^A},|{\phi}^{A'}\rangle, \ket{\phi^B}, |{\phi}^{B'}\rangle  \}$, takes the compact form  
\begin{equation}\label{Hphi}
H_{\phi} = \begin{pmatrix}
    0 & V_{pp\sigma} & V_{\rm{AB}} & 0 \\
   V_{pp\sigma}  & 0 & 0 & V_{\rm{AB}} \\
   V_{\rm{AB}} & 0 & 0 & V_{pp\sigma} \\
   0 & V_{\rm{AB}} & V_{pp\sigma} & 0
  \end{pmatrix},
\end{equation}
\noindent 
here the onsite energy of the $p$-orbitals of phosphorene was taken as the reference energy, and set to zero ($\varepsilon_{p}=0$).
Lastly, the matrix element $V_{\rm{AB}} = \langle \phi^{\rm{A}}|\hat{H}_{0}|\phi^{\rm{B}}\rangle = V_{\rm{A'B'}}$, which in terms of the Slater-Koster parameters reads, 
\begin{equation}\label{vab} 
V_{\rm{AB}} = m_{z}^{2}V_{pp\pi} - m_{y}^{2}\left( n_{y}^{2}V_{pp\sigma} + (1-n_{y}^{2})V_{pp\pi}\right),
\end{equation}
\noindent being $n_y$ is the cosine director along the $y$-axis as defined in Table\,\ref{SKdefinitions}, among the remaining Slater-Koster integrals. 

\subsection{Continuum Hamiltonian and low energy bandstructure}\label{ContHbands}

\begin{figure*}
\hspace{-0.5cm}
\includegraphics[width=12.0cm]{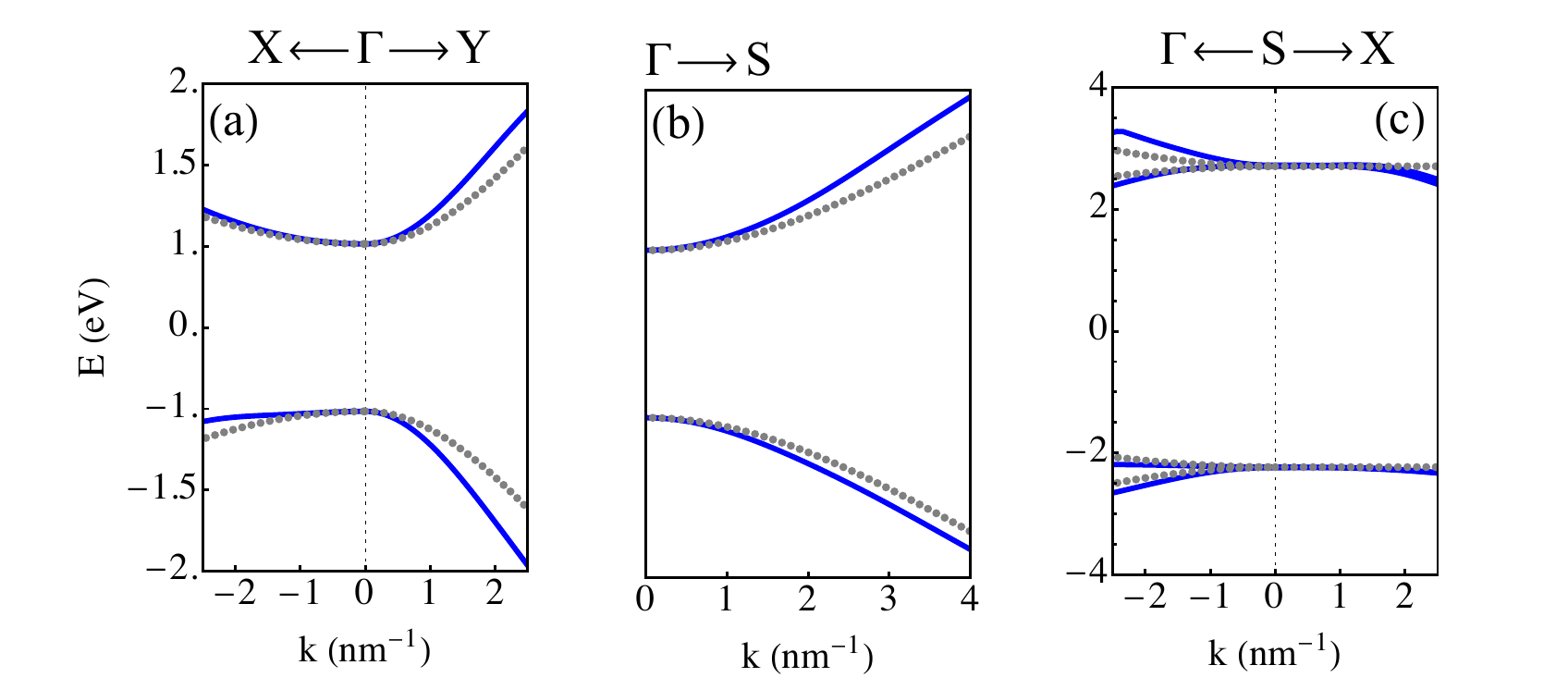}
    \caption{Electronic low energy bands for monolayer phosphorene without SOC along the paths ${\rm X} - \Gamma - {\rm Y}$ (a), $\Gamma - {\rm S}$ (b), and $\Gamma - {\rm S} - {\rm X}$ (c), respectively. The continuum (blue) lines  correspond to the DFT bandstructure calculations. The dotted (gray)
and squared-symbols (red) curves are obtained with the exact eigenvalues of the low energy effective model given in Eq. (8), using
 the SK-parameters provided by the DFT+W90 calculations, and the best fit parameters of DFT electronic bandstructure at low energies, respectively (see Table\,\ref{SKparam}). The Fermi energy of the DFT bands were shifted here to negative energies by half the band gap in order to facilitate their comparison with theoretical model. 
    }
    \label{NoSOC2}
\end{figure*}

Next, by Fourier transform the model of Eq.\,(\ref{Hphi}), an effective low energy $4\times4$ Hamiltonian in reciprocal space is obtained,
\begin{equation}\label{HphiR}
\small{
H_{\phi}(\bm{k}) = \begin{pmatrix}
 0 & g(\bm{k}) V_{pp\sigma} & f(\bm{k}) V_{\rm AB} & 0 \\
  g^{\ast}(\bm{k}) V_{pp\sigma} & 0 & 0 & f^{\ast}(\bm{k}) V_{\rm AB} \\
   f^{\ast}(\bm{k}) V_{\rm AB} & 0 & 0 & g^{\ast}(\bm{k}) V_{pp\sigma} \\
   0 & f(\bm{k}) V_{\rm AB} & g(\bm{k}) V_{pp\sigma} & 0    
\end{pmatrix},
}
\end{equation}
\noindent where  $\bm{k}=k_{x}\hat{x}+k_{y}\hat{y}$ is the wave vector lying in the plane of the phosphorene layer, and we have used the following definitions for the spectral functions,
\begin{equation}\label{spechtralf}
\small{
f(\bm{k}) = 2 e^{iyk_{y}} \cos(\frac{bk_{x}}{2}), \quad
g(\bm{k}) = e^{-ik_{y}h},
}
\end{equation}
with $b = 2 b_{1}\sin{\theta}$, $y=b_{1}\cos{\theta}$, $h = b_{2}\sin{\varphi}$, and $h+y = a/2$. The Hamiltonian matrix  Eq.\,(\ref{HphiR}) can be diagonalized exactly, however, in order to arrive to an effective low energy Hamiltonian we find it convenient to rewrite it in the basis $\{ \ket{\phi^A},|\phi^{B'}\rangle, |\phi^{A'}\rangle, |{\phi}^{B}\rangle  \} $ and perform a similarity transformation (Appendix\,\ref{CH}) that leads to a block-diagonal Hamiltonian composed of two $2\times 2$  matrix Hamiltonians ${\cal H}_{1}$ and ${\cal H}_{2}$ given by,

\begin{equation}
\label{Ho2X2}
\begin{split}
 {\cal H}_{\eta}(\bm{k}) 
&=(-1)^{\eta} \begin{pmatrix}
 \,\,\,\,\,\,\,\rm Re[{\cal V}_{\eta}(\bm{k})] & \,i\, \rm Im[{\cal V}_{\eta}(\bm{k})] \\
 -i\, \rm Im[{\cal V}_{\mu}(\bm{k})] &  -\rm \rm Re[{\cal V}_{\eta}(\bm{k})]
\end{pmatrix},
\end{split}
\end{equation}
\noindent 
where ${\cal V_{\eta}}(\bm{k})=V_{\rm{AB}}f(\bm{k})+(-1)^{\eta+1} V_{pp\sigma}g(\bm{k})$, and $\eta=1,2$. The lowest energy bands dispersion will correspond to the eigenvalues of (\ref{Ho2X2}) with $\eta=1$, 

\begin{equation}
\label{Eigen0}
\small
{ \cal E}_{\pm}  = \pm\sqrt{V^{2}_{pp\sigma} + 4 V_{\rm AB} V_{pp\sigma} \cos(k_{x}c)
 \cos(k_{y}d) +4 V^{2}_{\rm AB} \cos(k_{x}c)^2
}
\end{equation}
\noindent where the $\pm$ sign refers here to the electron/hole bands, being $c=b/2$ and $d=a/2$. The anisotropy of the generic bands is already evident within this model. Note also that the energy gap ${\cal E}_g$  at the $\Gamma$-point, is simply given by ${\cal E}_{+}(0)-{\cal E}_{-}(0)=2|V_{pp\sigma}+2V_{\rm AB}|$.  
\\

 \begin{table} 
\centering
\begin{tabular}{l c c c}
\hline \hline
  & $V_{pp\sigma}$ & \quad  $V_{pp\pi}$ & \quad  $V_{\rm{AB}}$ \\
  \hline
  DFT+W90 \quad \quad & 3.85 & \quad  -0.99 & \quad -1.02 \\
  Ref.\,[\onlinecite{Kurpas2016}] \footnote{Fitted to Eq.(\ref{Eigen0}) with DFT data of Ref.\,[\onlinecite{Kurpas2016}].}  \quad \quad & 3.15 & \quad -1.04 &  \quad-1.01 \\
  DFT$^*$  \footnote{Best fit to Eq.(\ref{Eigen0})  using our DFT raw data.}  \quad \quad & 3.30 & \quad -1.19 &  \quad-1.14 \\
  Ref.\,[\onlinecite{Menezes2018}]\,\,\, \quad \quad & 4.03 & \quad -1.14 & \quad-1.15 \\
\hline \hline
\end{tabular}
\caption{Numerical values of some Slater and Koster   parameters for phosphorene (in eV) without spin-orbit coupling, with the exception of  those from Ref.\,[\onlinecite{Kurpas2016}] where SOC was taken into account. The parameter $V_{\rm AB}$ was calculated using Eq.(\ref{vab}). Note that in all cases  $V_{pp\pi}\simeq V_{\rm AB}$.}
\label{SKparam}
\end{table}

As for the Slater-Koster parameters, they are extracted either, by fitting numerically the electronic band structure obtained by DFT calculations, or by a subsequent step of the DFT process by mapping  the electronic structure to a maximally localized Wannier basis of atomic orbitals using WANNIER90 (DFT+W90) \cite{Mostofi2014}. The values obtained here are listed in Table\,\ref{SKparam} together with these fitted to the DFT data of Ref.\,[\onlinecite{Kurpas2016}] and  SK-parameters reported in Ref.\,[\onlinecite{Menezes2018}] for sake of comparison. 

Starting from Eq. (\ref{HphiR}), we now focus on the resulting low energy $\bm{k}\cdot \bm{p}$-type effective Hamiltonian and its corresponding eigenvalues near the $\Gamma$-symmetry point. The expansion of the matrix elements around $\bm{k}=0$ of ${\cal H}_{\pm}({\bm k})$ up to quadratic terms in $k_x$ and $k_y$  leads to the simplified continuum Hamiltonian,
\begin{equation}
\label{Eigen1}
{\cal H}_{\Gamma}(\bm k) = (\alpha k_x^2+ \beta k_y^2 - \Delta)\tau_z -v_y k_y \tau_y \, ,
\end{equation}
where the coefficients in terms of the Slater-Koster parameters (using the result $V_{\rm{AB}}\simeq V_{pp\pi}$), are
\begin{equation}\label{Paramh2}
\begin{aligned}
v_y &= h V_{pp\sigma} - 2y V_{pp\pi}, \quad \text{(in units of $\hbar$)} \\
    \beta &=  \frac{h^2}{2} V_{pp\sigma} + y^2V_{pp\pi}, \\ \alpha & = \frac{b^2}{4} V_{pp\pi},
\end{aligned}
\end{equation}
\noindent with $\Delta={\cal E}_g/2$, and $\tau_y$, $\tau_z$ are the $y$- and $z-$Pauli matrices written in the electron-hole basis. The eigenvalues of Eq.\,(\ref{Eigen1}) produce the simplified formulas for the anisotropic dispersion of the conduction/valence bands,
\begin{equation}
\label{Eigen2}
 { \cal E}_{\Gamma_{\pm}} =\pm\sqrt{(\alpha k_x^2 +\beta k_y^2-\Delta)^2+v_y^2 k_y^2
 }.
\end{equation}

Next, we present useful simplified expressions for the low energy bands, as well as for the effective masses that are determined directly from Eq.\,(\ref{Eigen2}). They are up to second order in $k_x$ and $k_y$,
\begin{equation}
\label{EigenvG}
   { \cal E}_{\Gamma_{\pm}}^{(0)}(\bm k) =  \frac{\hbar^2}{2}\left(\frac{1}{m^*_{\pm,\Gamma X}} k^2_x + \frac{1}{m^*_{\pm,\Gamma Y}} k^2_y\right) \pm{\Delta},
\end{equation}
\noindent in which the Slater-Koster dependent effective masses along the $\Gamma-{\rm X}$  and $\Gamma-{\rm Y}$ path are given by 
\begin{equation}\label{Effmass}
\begin{split}
    \frac{1}{m^*_{\pm,\Gamma X}} &= \pm\frac{b^2}{2\hbar^2{\Delta}} \left( \abs{V_{pp\pi}V_{pp\sigma}}-2 V^2_{pp\pi} \right),\\
    \frac{1}{m^*_{\pm,\Gamma Y}}& =\pm\dfrac{a^2}{2\hbar^2{\Delta}} \abs{V_{pp\pi}V_{pp\sigma}}.
\end{split}
\end{equation}

\begin{table}
\centering
\vspace{0.15cm}
\begin{tabular}{lcccc}
\hline\hline 
    & $m^{*}_{+,\Gamma \rm X}$  \,\, & $m^{*}_{+,\Gamma \rm Y}$  \,\, & $m^{*}_{-,\Gamma \rm X}$  \,\, &  $m^{*}_{-,\Gamma \rm Y}$ \\ [1.0ex]
\hline 
    DFT+W90  
    &  $1.13$ &  $0.23$ & $-3.08$ & $-0.20$ \\
    Ref.\,[\onlinecite{LiY2014}]    &  $1.16$ &  $0.22$ & $-3.24$ & $-0.19$ \\
    Eq.\,(\ref{Effmass}) \footnote{Using the SK parameters of Ref.[\onlinecite{Menezes2018}] and Eq.(\ref{Effmass}).}   &  $1.18$ &  $0.17$ & $-1.18$ & $-0.17$ \\
    Ref.\,[\onlinecite{Kurpas2016}] \footnote{Reported in Ref.\,[\onlinecite{Kurpas2016}].}   &  $1.15$ &  $0.24$ & $-7.29$ & $-0.24$ \\
    Ref.\,[\onlinecite{LewYanVoon2015}]    &  $1.24$ &  $0.17$ & $-7.20$ & $-0.16$ \\
    Ref.\,[\onlinecite{Kurpas2016}] \footnote{Parabolic fit of the bands presented in Ref.\,[\onlinecite{Kurpas2016}].}    &  $1.28$ &  $0.42$ & $-4.06$ & $-0.44$ \\
\hline\hline  
\end{tabular}
\caption{\label{tableEM} List of the anisotropic conduction and hole effective masses for phosphorone monolayer (in units of the free electron mass) along the $\Gamma - \rm X $ and $\Gamma - \rm Y$ directions.} 
\vspace{-0.10cm}
\end{table}

Using the Slater-Koster parameters provided in Table\,\ref{SKparam} in the expressions above, we compute the phosphorene's effective masses around the $\Gamma$-point, 
and are listed together with those reported in the literature in Table\,\ref{tableEM}.  In Fig.\,\ref{NoSOC2} we present the lowest electronic bands of phosphorene at the vicinity of the $\Gamma$-  and $\rm S$-point in the absence of SOC effects. 
The blue (solid lines) correspond to the outcome of the first principles DFT calculations. The results from the minimal tight-binding model obtained through Eq.\,(\ref{Eigen0}) are plotted in squared-symbols (red) by using the best fit to the SK parameters extracted from the DFT data, whiles the dotted (gray) curves correspond to the evaluation of Eq.\,(\ref{Eigen0}) using the SK parameters obtained through the DFT+W90 procedure. Interestingly, despite its  simplicity, our minimal tight-binding model  clearly displays the  expected asymmetry of the electronic bands around the high symmetry points characteristic of phosphorene band structure\cite{Rodin2014, Popovic2015, Kurpas2016}. 

In overall, we find that the minimal base tight-binding model reproduces accurately the lowest bands, though showing a slight deviation for the valence band in the $\rm{\Gamma - X}$ direction where, in comparison with the DFT bands, a smaller effective mass is obtained. Such slight discrepancy is not actually owed that we have neglected the $p_x$ and $s$ orbitals at the present stage of the tight-binding model. It is shown {{\it a posteriori}} that taking into account such orbitals, does not lead to any appreciable change of the curvature of the bands around the $\Gamma$-point.

\section{Tight Binding model with spin-orbit interaction}\label{TBsoc}

We now proceed to examine the relativistic effects in monolayer phosphorene. We focus on the spin-orbit interaction of the intrinsic (atomic) and extrinsic (Rashba) type. The former arises naturally due the local  spin-orbit interaction of the atomic outer-shell $3s$ and $3p$-electrons owing the presence of the potential $V({\bm r})$ of the phosphorous ions. The latter (Rashba-type) can be present owing the Stark effect, and it is produced by an external uniform electric field $\bm E$ perpendicular to the phosphorene plane, breaking the phosphorene spatial inversion symmetry along such direction.

The atomic spin-orbit interaction is modeled with the Hamiltonian,
\begin{equation}
\label{Hso}
    {H}_{so}= \frac{\hbar}{2m^2c^2}[\nabla V({\bm r})\times {\bm p}]\cdot {\bm s} =\xi(r) {\bm L}\cdot {\bm S}
\end{equation}
\noindent where $m$ is the free-electron mass, $c$ is the speed of light in vacuum, ${\bm p}$ is the linear momentum operator, ${\bm s}=(s_x,s_y,s_z)$, is the vector of the Pauli matrices, ${\bm S} = \frac{\hbar}{2}{\bm s}$ denotes the electron spin vector operator, and ${\bm L}$ is the orbital angular momentum operator. Within the two-center approximation, the atomic potential is assumed to be spherically symmetric $V(\bm{r})\rightarrow V(r)$, therefore the function $\xi(r)=\frac{1}{m^2c^2}\frac{1}{r}\frac{\partial V}{\partial r}$ carries all the radial dependence. In addition, the spin-orbit interactions between neighboring phosphorous atoms, and beyond are assumed to be negligible. Hence, only the on-site spin-orbit interaction terms shall be considered in our approach. 

Regarding the Stark effect, we include it by introducing an external electric field ${\bm E}=E_z\hat z$ perpendicular to the phosphorene plane. This field may arise from various sources such as an externally applied voltage or charge impurities at the substrate interface. We model this effect through the dipole term in the Hamiltonian,
\begin{equation}\label{HSE}
{H}_{\rm SE}= -eE_z\,z, 
\end{equation}
where $e$ is the charge of the electron.
As this interaction yields a spatial symmetry breaking in the $\hat{z}$ direction, thus within tight-binding theory, the Stark effect leads to on-site transitions between orbitals with opposite parity in the direction of the field. Thus, as a consequence of the spatial inversion symmetry breaking in the phosphorene layer, a sizeable spin-splitting can be exhibited in the electronic bands. 

Here the net effect of the spin-orbit interaction and the Stark effect are  included into the model by using an extended spin-dependent basis $|\Psi\rangle=|\psi\rangle\otimes \{\uparrow,\downarrow\}$.  In such basis
 $|\psi\rangle=\{|\phi\rangle,|\chi\rangle\}$, in which $|\phi\rangle$ are described in Eq.(\ref{basisphi}), whiles  $|\chi\rangle$
takes into account the hybridization of the in-plane $p_x$ and $p_y$-like orbitals, and the $s$-like orbitals, all within the same footing. Explicitly $|\chi\rangle=\{|\Theta^{\rm{A}}\rangle, |\Theta^{\rm{A'}}\rangle, |\Theta^{\rm{B}}\rangle, |\Theta^{\rm{B'}}\rangle, |s^{\rm{A}}\rangle, |s^{\rm{A'}}\rangle, |s^{\rm{B}}\rangle,|s^{\rm{B'}}\rangle\}$, being
\begin{equation} \label{basistheta}
    \begin{split}
       |\Theta^{\rm A/A'}\rangle&=\,\,\,\,\,{n_{lx}}|{p_{x}}^{\rm A/A'}\rangle \pm {n_{y}}|{p_{y}}^{\rm A/A'}\rangle, \\
       |\Theta^{\rm B/B'}\rangle&= -n_{lx} | {p_{x}}^{\rm B/B'}\rangle \mp n_{y}|{p_{y}}^{\rm B/B'}\rangle. 
    \end{split}
\end{equation} 
with $n_{lx}=(-1)^l \sin{\theta}$, and $n_y=\cos{\theta}$, where $l=1,2$ and distinguishes the first two in-plane neighbors ${\rm B(B')}$ and ${\rm A(A')}$ atoms (see Fig.\,\ref{NoSOC1}(b)). 

We then can construct a full spin-dependent  $24\times24$ Hamiltonian in the basis $|\Psi\rangle$. It reads,
${\cal H}_{T}={\cal H}_{0}+{\cal H}_{SO}+{\cal H}_{SE}$, where
\begin{equation}\label{CompleteH}
    \mathcal{H}_{0}=
    \begin{pmatrix}
    \cal{H}_{\phi} & \mathcal{U} \\
   \mathcal{U}^{\dagger}  & {\cal H}_{\Theta,s} 
  \end{pmatrix},
\end{equation}
with $\mathcal{H}_{\phi}=H_{\phi} \otimes \{\uparrow,\downarrow\}$ and $\mathcal{H}_{\Theta,s} =H_{\Theta,s} \otimes \{\uparrow,\downarrow\}$, in which $H_{\Theta,s}$ contains the on-site and hopping integrals between the $\{|\Theta\rangle \}$ and $\{|s\rangle \}$ states, whereas $\mathcal{U}$ is the coupling matrix between the $\{|{\phi}\rangle\}$ and $\{|\Theta\rangle,|s\rangle\}$ states. The Hamiltonians ${\cal H}_{SO}$ and ${\cal H}_{SE}$ correspond to the spin-orbit and Stark Hamiltonians written in the basis $|\Psi\rangle$, respectively. Details of the derivation 
a low energy effective  Hamiltonian starting from ${\cal H}_{T}$ using a L\"owding transformation method is provided in the Appendix \ref{SOC}.

It is noteworthy that the matrix elements of the spin-orbit sector of the total Hamiltonian ${\cal H}_{T}$ have the general form
\begin{equation}
\langle \psi^{(u)}_{\nu,\sigma}| \xi(r){\bm L}\cdot {\bm S} |\psi^{(u)}_{\nu',\sigma'}\rangle 
\rightarrow \Gamma^{\sigma,\sigma'}_{\nu,\nu'}(u)
\end{equation}
\noindent with $(\nu,\nu')=\{s,p_x,p_y,p_z\}$ denoting the character of the atomic orbital of a given phosphorous atom $u=\{A,A',B,B'\}$ of the primitive cell, and $(\sigma,\sigma')=\{\uparrow,\downarrow\}$ describing its spin state. Operationally, in order to arrive to such matrix elements, the states  $|\psi^{(u)}_{\nu,\sigma}\rangle$ are written in the basis of the eigenkets of the 
angular momentum $|\ell, m,\sigma\rangle$, with $\ell = 0,1$ and $m = 0, \pm 1$, the angular and magnetic quantum numbers.   
Hence, matrix elements of the type 
$\xi_{\ell} \langle \ell,m,\sigma|{\bm L}\cdot {\bm S}|\ell',m',\sigma'\rangle\delta_{\ell,\ell'}$ can  be straightforwardly calculated, where here the parameter $\xi_{\ell}$ measures the strength of the spin-orbit coupling.  Such parameter is given by the integration of the radial part of the $\ell$-orbitals of phosphorus atoms; here all the nonvanishing matrix elements are those for $\ell=1$. As this parameter cannot be known within the tight-binding approach, we rely on the results obtained by first principles calculations after Wannierization of the electronic bandstructure.

However, althought it is indeed possible to introduce explicitly all the $\Gamma^{\uparrow,\downarrow}_{\nu,\nu'}(\mu)$ in the model, the resulting Hamiltonian expressions are rather complex, rendering them far less illuminating. Therefore, we consider instead its mean absolute value within perturbation theory, defined hereafter by $\xi_p$. From the DFT+W90 caculculations we extract  $\xi_p= 0.047$ eV for monolayer phosphorene. Such value is slightly greater than the reported for silicene ($\xi_p \approx 0.034$ eV)\cite{Liu2011}  as expected, owing that the atomic numbers of silicon and phosphorus differ just by one. On the other hand, the Stark effect leads to on-site transitions between $s$ and $p_z$ orbitals of the type $z_{sp}=\langle s|z|p_z\rangle$.
Thus the energy associated to the Stark effect is $\Delta_{z} = -E_z e z_{sp}$, where the magnitude of the electric field, $E_z$, ranges typically between $1$ and $5$V/nm \cite{Kurpas2016}. As for $z_{sp}$, it depends on the atomic size,  here we estimate $z_{sp}\approx 4.5 a_0$, where $a_0=0.0529$ nm is the Bohr radius.   

\subsection{Hamiltonian near the $\Gamma$ point with SOC}

Once we have considered the presence of the Stark effect and intrinsic spin-orbit coupling in phosphorene, we then proceed to determine the effective continuum Hamiltonian written in reciprocal space near the $\Gamma$-point and valid for its low energy quasi-particles (see Appendix E). The Hamiltonian reads, 
\begin{equation}\label{HGSOC}
\begin{aligned}
{H}(\bm k) = {} &  (\alpha k_x^2+\beta k_y^2  -\Delta)s_{\rm o}\tau_z - v_y k_y s_{\rm o}\tau_y\\
 &+\lambda_{Rx} k_x s_y\tau_z -\lambda_{Ry} k_y s_x\tau_z + \Delta_{R_{0}} s_x\tau_y ,
\end{aligned}
\end{equation}
where, $(\tau_y,\tau_z)$  and $(s_x,s_y)$ are the usual Pauli matrices acting on the electron-hole bands, and on the physical spin, respectively, being $s_{\rm o}$ the unit matrix in spin space. The first two terms in Eq.(\ref{HGSOC}) characterize the low-energy quasiparticles in the absence of SOC effects already discussed in the previous section. The next two terms proportional to $\lambda_{Rx}$ along the $\Gamma-{\rm X}$, and to $\lambda_{Ry}$ along the $\Gamma-{\rm Y}$ direction, describe an anisotropic $\bm{k}$-linear Rashba-SOC arisen because of the Stark effect. The last term accounts for a spin-orbit dependent interaction between conduction and hole bands with strength $\Delta_{R_0}$. It is also of Rashba nature, though $\bm{k}$-independent, as it is proportional to the external electric field. 

In terms of the Slater-Koster overlapping integrals, the  Rashba parameters are found to be (see Appendix\,\ref{SOCRG})
\begin{equation}\label{LambdaRx}
\begin{split}
\lambda_{Rx} & \simeq 4 b_1 \zeta_0  {(\sin{\theta})}^2 \\
\lambda_{Ry} & \simeq  2 b_2 \zeta_0 {(\cos{\theta}\sin{\varphi})}^2  \\
\Delta_{R_0} & = \frac{\lambda_{Ry}}{b_2 \sin{\varphi} }\, , 
\end{split}
\end{equation}
\noindent where
\begin{equation}
\label{zeta0}
\zeta_0 = \frac{ \Delta_{z} \xi_p V_{ss\sigma}V_{sp\sigma} \cos{\varphi} }{(V_{\rm AA} - V_{pp\sigma})(\varepsilon_s -\varepsilon_p)^2}, 
\end{equation}
\noindent and $ V_{\rm{AA}} =n^2_{lx}V_{pp\pi} - n^2_y (m^2_y V_{pp\sigma} + (1 - m^2_y) V_{pp\pi})$. Here,  $\lambda_{Rx}$, $\lambda_{Ry}$, and $\Delta_{R_0}$ are all proportional to the magnitude of the electric field $E_z$ via $\Delta_{z}$, as well as  with the intrinsic spin-orbit parameter associated to $p$-orbitals of phosphorene $\xi_p$. Also, as can be seen in Eq.\eqref{zeta0}, they depend linearly on the onsite hybridization between $s$ orbitals $ (V_{ss\sigma})$ and between $s$ and $p$ orbitals ($V_{sp\sigma}$) of the phosphorous atoms.  The strong anisotropy of the Rashba spin-splitting of the bands can be estimated through the ratio 
\begin{equation}
\frac{\lambda_{Rx}}{\lambda_{Ry}}=\frac{2b_1}{b_2}\frac{\tan^2{\theta}}{\sin^2{\varphi}}\simeq 18.7,
\end{equation}
\noindent which incidentally, at first order,  is not related to the SK-parameters,  depending only upon geometrical factors of the phosphorene atomic configuration. Using the SK-paramaters obtained within the  DFT+W90 procedure,  we find that at a typical electric field of $E_z=3$\,eV/nm, the  linear in $\bm{k}$ Rashba coupling coefficients are $\lambda_{Rx}=0.2576$\,meV\,nm and $\lambda_{Ry}=0.01379$\,meV\,nm, respectively, whereas the estimates as such field for the term independent of $\bm{k}$ gives us $\Delta_{R_0}\simeq 0.168\,$meV. Note that within this model, the intrinsic SOC does not appear explicitly here,  as it turn out to be vanishing smaller. Therefore, we can safely argue that is the Rashba coupling type that gives rise to the dominant spin-orbit interaction at the $\Gamma$-symmetry point of phosphorene, being the $\bm{k}$-linear dependent term  the most dominant at typical fields.  

Neglecting the spin-dependent interaction term $\Delta_{R_0}s_x \tau_y$ between
conduction and valence bands, 
the eigenvalues of the Hamiltonian in Eq.\,\eqref{HGSOC} give the following exact band dispersion laws,
\begin{equation}
\label{eigenSOCsimp}
        E^{(\mu)}_{\sigma}(\bm k) = \mu\sqrt{v^2_y k^2_y + (\alpha k^2_x + \beta k^2_y - {\Delta} +\sigma \kappa)^2}, 
\end{equation}
where $\mu=\pm$ describes the conduction/valence bands, $\sigma =\pm$ denotes the spin-state for a given band, and $\kappa=|\bm \kappa|$, with $\bm \kappa =(\lambda_{Rx} k_x,\lambda_{Ry} k_y)$, such that  $\kappa \exp{\pm i \phi_k} = \lambda_{Rx} k_x \pm i \lambda_{Ry} k_y$, with $\tan\phi_k = \lambda_{Ry} k_y/(\lambda_{Rx} k_x)$. 

\begin{figure}
    \includegraphics[width=8.5cm]{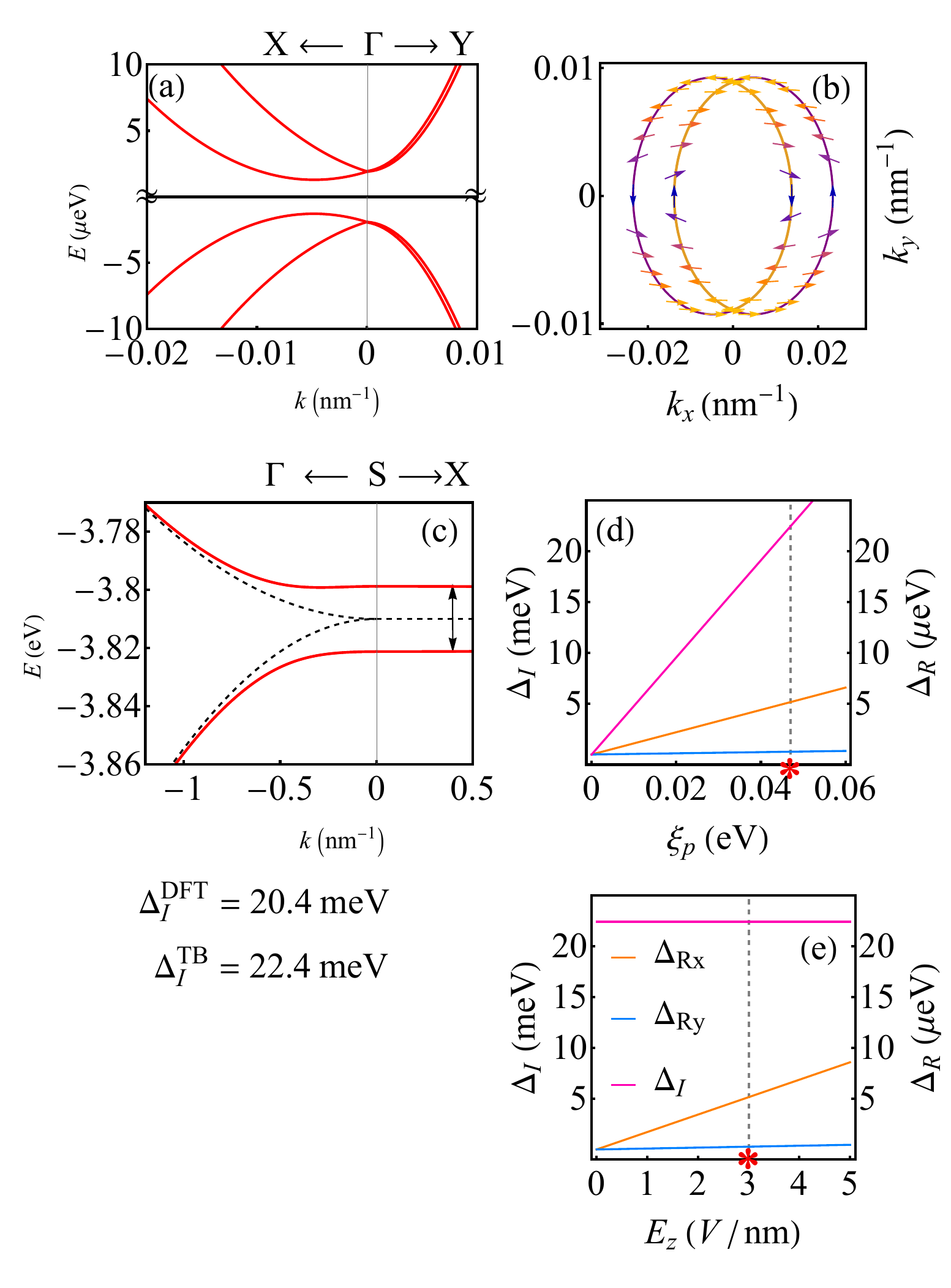}
    \vspace{-0.1cm}
    \caption{Band structure of phosphorene with SOC. In (a) the asymmetric spin splitting of the bands between the directions $\Gamma - {\rm X}$ and $\Gamma - {\rm Y}$, giving a ratio $\lambda_{Rx}/\lambda_{Ry} = 18.7$.
    (b) Depiction of the spin texture of the valence bands 
    for an energy of $10\mu$eV. The plot in (c) shows the intrinsic spin-orbit splitting at the $\rm S$-point between the valence bands $\Delta^{\rm TB}_{I}=22.4$meV, red(continuum) curves. The black(dashed) curves correspond to the same bands without SOC. The splitting obtained from the DFT calculations is shown for comparison $\Delta^{\rm DFT}_{I} = 20.4$meV.  For (a)-(c) the electric field is $E_z=3$V/nm and the intrinsic spin-orbit parameter was set to $\xi_p=0.047$eV. Panels (d)-(e) shows the dependence of the intrinsic SOC splitting at the ${\rm S}$ point denoted by $\Delta^{\rm TB}_{\rm I} = \Delta_{\rm I}$ and the Rashba spin-splitting in the $k_x$ and $k_y$ directions, $\Delta_{Rx}$ and $\Delta_{Ry}$, respectively, as a function of  $\xi_p$ for a fixed electric field of $E_z=3$V/nm (d), and (e) as a function of the electric field $E_z$ for  $\xi_p=0.047$\,eV. The Rashba splittings $\Delta_{Rx}$ and $\Delta_{Ry}$, were calculated for $k_F=0.01\,nm^{-1}$, where $k_F$ is the Fermi wave number. 
    }.
    \label{BandsSOCN}
\end{figure}

In Fig.\,\ref{BandsSOCN}(a) we present plots of the conduction and valence bands using expression (\ref{eigenSOCsimp}) showing  the large anisotropy of the band spin-splittings ocurring near the $\Gamma$-point when an electric field of $E_z=3$\,eV/nm is considered. If we define the energy spin-splitting of the bands at a given Fermi wave number $k_F$ as $\Delta_{R_i}(k_F)=2\lambda_{R_{i}} k_F$ with $i=x,y$ and choosing a small $k_F=0.01$\,${\rm nm}^{-1}$ (for a small carrier density), yields spin-splitting energies of $\Delta_{Rx}(k_F)=5.15$\,$\mu$eV and $\Delta_{Ry}(k_F)=0.28$\,$\mu$eV, respectively (Fig.\ref{BandsSOCN}(a), (d), and (e)). Whiles for higher carrier densities ($n_F=1\times10^{12}$\, ${\rm cm}^{-2}$) entailing a Fermi wave number of about  $k_F=0.25$\,${\rm nm}^{-1}$, we get spin-splitting energies of $\Delta_{Rx}(k_F)=0.13$\,meV and $\Delta_{Ry}(k_F)=6.89$\,$\mu$eV, respectively. The monotonic linear behavior of the Rashba spin-splitting with the SOC parameter $\xi_p$ and with the field $E_z$ is depicted in Fig.3(d)-(e). The plot in Fig.3(c) shows the valence bands in the vecinity of the the $\rm S$-point, with (red-continuos curves) and without (black dashed-lines) intrinsic SOC. More details of the instrinsic SOC are discussed in the following subsection.  

The corresponding normalized eigenvectors of Eq.(\ref{HGSOC}) for the conduction $(c)$ and valence bands $(v)$ read,
\begin{equation}
\begin{split}
|\psi^{(c)}_{\sigma}(\bm k)\rangle &= \frac{1}{\sqrt{2}} \begin{pmatrix}
 \sigma e^{-i \phi_{k}} \cos(\vartheta_{\sigma}/2) \\
 i \cos(\vartheta_{\sigma}/2) \\
 -\sigma i e^{-i \phi_{k}} \sin(\vartheta_{\sigma}/2) \\
 \sin(\vartheta_{\sigma}/2) \\
\end{pmatrix}, \\
|\psi^{(v)}_{\sigma}(\bm k)\rangle &= \frac{1}{\sqrt{2}} \begin{pmatrix}
 \sigma e^{-i \phi_{k}} \sin(\vartheta_{\sigma}/2) \\
  -i \sin(\vartheta_{\sigma}/2) \\
 \sigma i e^{-i \phi_{k}} \cos(\vartheta_{\sigma}/2) \\
  \cos(\vartheta_{\sigma}/2)
\end{pmatrix},
\end{split}
\end{equation}
where we have defined $  \vartheta_{\sigma} = \arctan (v_y k_y / {E}_{\sigma})$, 
and we have omitted the superscript ($c/v$) in ${E}_{\sigma}$ to easy the notation.   The spin-texture is then readily calculated,
\begin{equation}
\begin{aligned}
    \langle {\bm S} \rangle_{\mu,\sigma} & = \mu \sigma\Big(-\frac{\lambda_{Ry} k_y}{\kappa}, \frac{\lambda_{Rx} k_x}{\kappa}, 0\Big)\\
    &=\frac{\mu \sigma}{\kappa}({\hat z}\times \bm \kappa).
\end{aligned}
\end{equation}
Observe that the relative magnitude of the spin-orientation vectors depend on the ratio $\lambda_{Rx}/\lambda_{Ry}$ through $\bm \kappa$, and that Rashba coupling anisotropy in phosphorene imposes $\langle {\bm S} \rangle_{\mu,\sigma}\cdot \bm k\neq 0$ in general, contrasting with what it occurs in graphene with Rashba spin-orbit coupling, where $\langle {\bm S} \rangle_{\mu,\sigma}\cdot \bm k= 0$ is always satisfied. In Figure\,\ref{BandsSOCN}(b) we show the spin texture around the Fermi energy contour formed by the valence bands for an energy of $-10\,\mu$eV, an electric field $E_z=3$\,V/nm, and a spin-orbit parameter $\xi_p=0.047$\,eV. The spin splitting asymmetry near the $\Gamma$ point is evident, having a non-tangential direction of the spin with respect to the energy Fermi contour, except in the points $(k_x^F,0)$ and $(0,k_y^F)$. We also explored the range of the linear Rashba spin-splitting $\Delta_{R_x}$ and $\Delta_{R_y}$ for a fixed $\xi_p=0.047$\,meV as a function of the external field $E_z$ (Fig.3(e)). The counterparts for a fixed field $E_z =3$\,V/nm as a function of $\xi_p$ are plotted in (Fig.3(d)).

\begin{figure}
\centering
    \includegraphics[width=8.5cm]{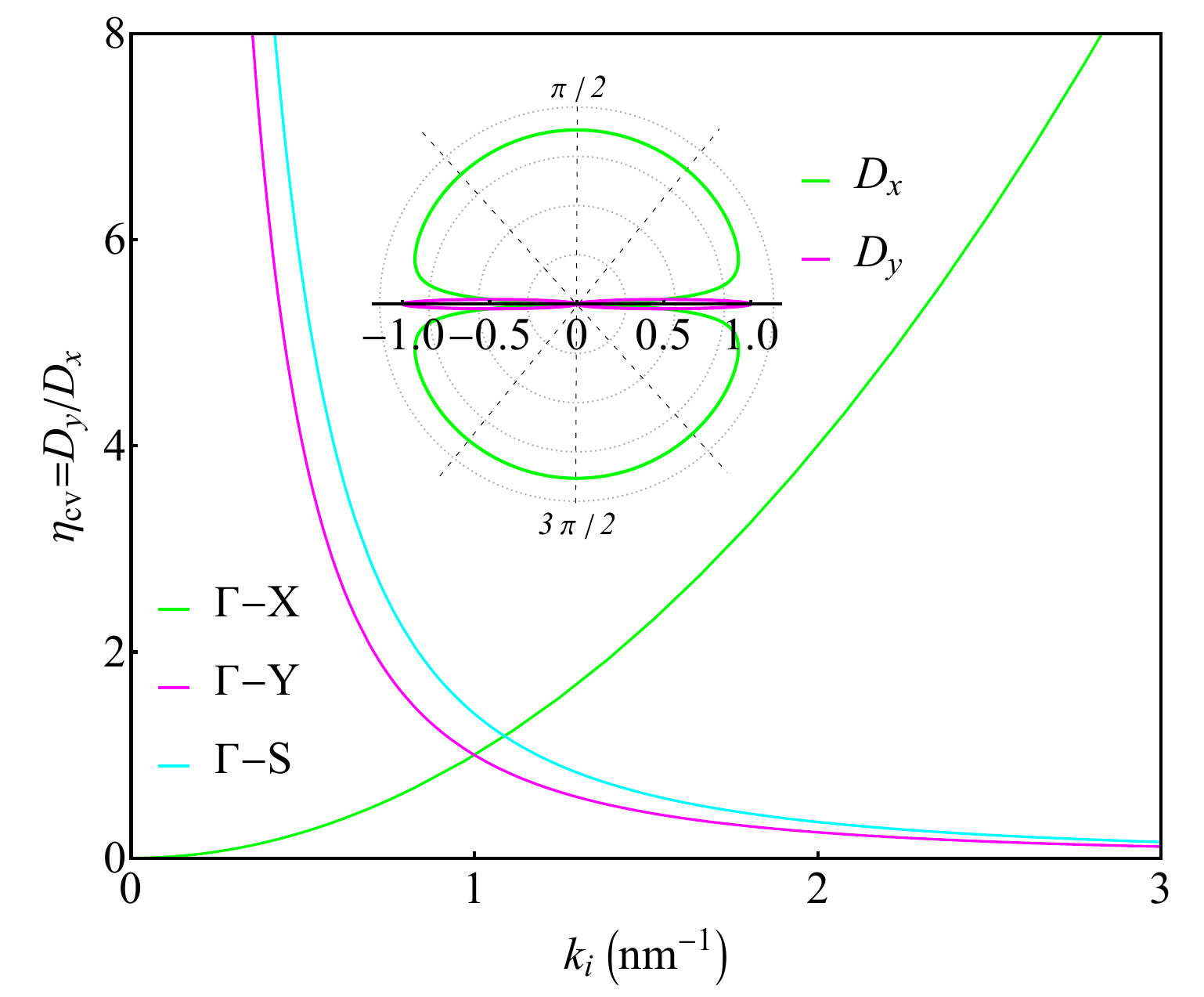}
    \vspace{-0.2cm}
    \caption{ Dipole strength ratio $\eta_{cv} = D_y/D_x$ as a function of $k_i$ in the $\Gamma-{\rm X}$ ($i=x$, green), the $\Gamma-{\rm Y}$ ($i=y$, magenta), and the $\Gamma-{\rm S}$ ($i=s$, cyan) directions. The inset shows the dependence of the dipole strength $D_x$, and $D_y$ with the angle $\phi_k$, exhibiting the strong anisotropy.}
    \label{dipole}    
\end{figure}

It is also illustrative to calculate dipole strength coupling between the conduction and valence band along the $x$ and $y$ direction as a function of $\bm k$, defined here as
\begin{equation}
{  D}_i(\bm k) =\sum_{\sigma}|\langle \psi_{\sigma}^{(c)}|\frac{m_o}{\hbar}\frac{\partial H(\bm k)}{\partial k_i}|\psi_{\sigma}^{(v)}\rangle|^2\, , \quad i=x,y
\end{equation}
\noindent with $m_o$ the free electron mass. This leads to dipole strength anisotropy 
\begin{equation}
\begin{split}
{  D}_x(\bm k) &=\lambda_{Rx}^2\sin(\frac{\vartheta_{+}+\vartheta_{-}}{2})^2 \sin{\phi_k}^2,\\
{  D}_y(\bm k) &=\lambda_{Ry}^2\sin(\frac{\vartheta_{+}+\vartheta_{-}}{2})^2 \cos{\phi_k}^2
\end{split}
\end{equation}
\noindent in units of $m_o^2/\hbar^2$. Interestingly, the ratio of the dipole strengths $\eta_{cv}={  D}_y/{  D}_x$ along the $x$ and $y$ direction does not depend on the spin-orbit coupling strengths, nor in its anisotropy, and gives the simple formula $\eta_{cv}= k_x^2/k_y^2$. Plots of the dipole strength along different directions are depicted in Fig.\ref{dipole}. 

\subsection{Hamiltonian around the $\rm{S}$ point with SOC}

Using L\"owdin transformation theory  it can be shown (appendix \ref{SOC}) that in the vecinity of the $\rm S$-symmetry point ${\bm k}_s =(k_{sx},k_{sy})$, the full $8\times8$ spin-dependent Hamiltonian  is given by
\begin{equation}
\label{HwithISOC}
 {\cal H}({\bm k}) = {\cal H}^{\prime}({\bm k}) s_o  + {\cal H}_{I}({\bm k})
\end{equation}
with ${\bm k}= {\bm k}_s +{\bm q}$, being ${\bm q} =(\pi/b,\pi/a)$ the location of the  $\rm S$-point in the rectangular Brillouin zone referred to zone center $\Gamma$. 
Here,
\begin{equation}
\label{H4X4}
{\cal H}^{\prime}= \begin{pmatrix} 
{\cal H}_{2} & 0 \\
0 & {\cal H}_{1} 
\end{pmatrix},
\end{equation}

\noindent where $ {\cal H}_{1,2}(\bm k)$ were already described in  Eq.(\ref{Ho2X2}), and comprises the diagonal sectors in spin of the full Hamiltonian. 
In Eq.(\ref{HwithISOC}) ${\cal H}_{I}$ arises as the dominant spin-orbit interacting term at this symmetry point, and it turn out to be of purely  intrinsic nature, as it is owed to the atomic SOC of the outer shell electrons with the nucleous of the phosphorous atoms. It has the compact form
\begin{equation}
{\cal H}_I({\bm k}) =  
\lambda_I {\rm Im} [h(\bm{k})]\tau_z \sigma_y s_o- \lambda_I {\rm Re} [h(\bm{k})]\tau_z \sigma_z s_x
\end{equation}

\noindent where $ h(\bm k) = 2\sin(\frac{bk_{x}}{2}) e^{ik_{y} a/2}$, and we have introduced the new the Pauli matrices $(\sigma_y, \sigma_z)$ on order to describe the coupling of the otherwise (in the absence of SOC) double degeneracy (four degeneracy taking into account the spin) of the conduction and valence bands along the $\rm S$-$\rm X$ direction. The constant $\lambda_{I}$ modulates the strength of the intrinsic spin-orbit interaction, and as a function of the Slater-Koster parameters reads,
\begin{equation} \label{DeltaS}
    \begin{split}
       \lambda_{I} &= \frac{2 V_{pp\sigma} \xi_p  \cos\theta \sin{\theta}\sin^2{\varphi}} {V_{\rm{AA}} - V_{pp\sigma}},
    \end{split}
\end{equation}
note that it depends only on the intrinsic spin-orbit parameter $\xi_p$, and on the hybridization between $\sigma$- and $p$-orbitals of phosphorene, whiles those involving $V_{ss\sigma}$ and $V_{sp\sigma}$ for instance, do not play any role here, and as a consequence, in the spin-splitting of the bands.

In the following, to easy the notation, we shall make the replacement ${\bm k}_s\rightarrow {\bm k}$, with the assumption that the shifting with zone center has been already performed. An effective  $\bm{k}\cdot \bm{p}$-type low energy Hamiltonian can be derived by expanding  Eq.(\ref{HwithISOC}) up to second order in ${\bm k}$, however such  Hamiltonian turns out to be cumbersome and not too illuminating to describe it here.
On the other hand, it is compelling that given the smallness of  $\lambda_I$ we can  consider only the leading order expansion of ${\cal H}_I$. This renders the full Hamiltonian  exactly diagonalizable, yielding to the following eigenvalues for the (two) conduction bands at the $\rm S$ point,

\begin{widetext}

\begin{equation}\label{EivaloS}
{\cal E}^{\sigma}_{S,{\nu}}(\bm k)=\sqrt{V^2_{pp\sigma} +4 \lambda^2_I+4 V^2_{\rm AB}\sin^2{\Big(\frac{b k_{x}}{2}}\Big)+4\sigma \nu V_{pp\sigma} \sqrt{V^2_{\rm AB}\sin^2{\Big(\frac{b k_{x}}{2}}\Big)\sin^2{\Big(\frac{a k_{y}}{2}}\Big) +  \lambda^2_I }},
\end{equation}

with $\sigma=\pm$ for a given spin state, and   $\nu=\pm $ for the conduction band 1 or 2, respectively. As the electron-hole symmetry is not broken within this model, we have similarly  $-{\cal E}^{\sigma}_{S,\nu}(\bm k)$ for the valence bands.  
\end{widetext}

From the expression (\ref{EivaloS}) we can extract the energy spin-splitting at ${\bm k}=0$ of a given conduction or valence  band ($\nu$), and it gives simply,

\begin{equation}
\Delta_I^{TB} = {\cal E}^{\sigma}_{S,{\nu}}(0)-{\cal E}^{-\sigma}_{S,{\nu}}(0) = 4 |\lambda_I|.
\end{equation}

According to this formula, using (\ref{DeltaS}) and the output of the  tight-binding SK parameters obtained through Wannieralization of the bandstructure, the SOC spin-splitting at the $\rm S$-point between the conduction/valence bands is $\Delta^{\rm TB}_{I}=22.4$\,meV, which is in very good agreement with the value obtained directly by our DFT calculations ($\Delta^{\rm DFT}_{I} = 20.4$\,meV). Note that for phosphorene $\lambda_I^{TB}=5.6$\, meV is greater than the value estimated for Silicene ($\lambda_I=3.9$\, meV) but much smaller than the one calculated for Germanene ($\lambda_I=43$\,meV)\cite{Liu2011} at the Dirac $\rm K$-point.

\section{Conclusions and summary}

We have introduced a minimal Slater-Koster multi-orbital tight-binding model amenable for description of the low energy quasi-particles in two-dimensional phosphorene under spin-orbit effects.  Our theoretical analysis  was  complemented with numerical DFT calculations of the electronic band structure with a {\it posteriori} construction of a Hamiltonian written in a suitable maximally localized Wannier basis employing the Wannier90 code\cite{Mostofi2014}. The latter allow us to extract the relevant Slater-Koster hopping integrals that were subsequently utilized in our effective models. Our analytical tight-binding approach yields useful expressions for the effective $\bm{k}\cdot \bm{p}$-type Hamiltonians describing the large anisotropy of the electronic band structure of monolayer phosphorene near the Fermi energy at the high symmetry points. The model predicts also a strong anisotropy of the extrinsic Rashba SOC when a perpendicular electric field is applied. It was found that the dominant spin-orbit coupling  
effects at the zone center $\Gamma$ is of the extrinsic (field dependent) Rashba-type, comprising two main terms, one which is linear in momentum, and other one that is momentum-independent, though rather weak.   In contrast, at the $\rm S$ symmetry point, is the intrinsic SOC which is the leading interaction, being the Rashba-type interaction totally absent at this point. 

We found that anisotropy of the Rashba coefficients follows the rule $\lambda_{R_x} \simeq 19 \lambda_{R_y}$, independent of the field. The tunability of the Rashba coefficients could lead to sizable spin-splittings. It is shown that a typical fields of $\sim3$\,V/nm and carrier densities of the order of $10^{12}$\,$cm^{-2}$ the  
energy spin-splitting of the bands in the $\Gamma$-X direction is in the tenths of meVs ($\sim 0.13$\,meV), whiles it ranges in the $\mu$eVs in the $\Gamma$-Y path ($\sim 7 \mu$eV).  It should be mentioned that an enhancement of about one order of magnitude of the Rashba coefficients can be achieved in phosphorene without the need of an external electric field, but by means of proximity effects with suitable substrates, as very recent has been reported in phosphorene/WSe$_2$ bilayer\cite{MMFabian2023}.

In contrast with what occurs near the $\Gamma$-symmetry point, our tight-binding model also predicts that at the $\rm{S}$-symmetry point, is rather the (atomic) intrinsic type of spin-orbit coupling which turn to be the dominant interaction, whiles the Rashba-type is basically negligible at this point. The latter is  in agreement with our DFT calculations. We found that the energy spin-splitting developed at the $\rm S$-symmetry point is about $22 $\,meV. Given the ranges of energies around the $\rm{S}$ point, the results presented here can provide valuable insights into the physics of electron-hole recombination processes in phosphorene, particularly through direct and indirect optical transitions, such as thermal excitations (electron-phonon interactions). In this context, our effective Hamiltonian with SOC around the $\rm{S}$ point would be useful for example for the study of anisotropic thermal conductivity in phosphorene.

Lastly, simple formulas for the inter-band dipole-strengths are presented, revealing the origin of the strong anisotropic behavior of the low energy bands in phosphorene. We hope that this studies and the effective Hamiltonian models with spin-orbit effects introduced here could serve as a good starting point for further scrutiny of the spin-dependent transport as well as its opto-spintronic properties in phosphorene and possibly in its nanoribbons.

\section{Acknowledgements}
F.M. and M.P. acknowledge funding from PAPIIT-DGAPA-UNAM through projects IN113920 and IN111624. D.F. is grateful to the support received from Yachay Tech University. 

\appendix

\section{Phosphorene's crystalline structure and geometrical parameters}\label{structure}
Single layer (black) phosphorene possesses a two-dimensional (2D) honeycomb puckered structure (see Fig.\,\ref{NoSOC1}(a-b)) of phosphorus atoms covalently bounded to their three nearest neighbors throughout a $sp^3$ orbital hybridization. 
It has a rectangular 2D primitive lattice with a basis formed by four atoms sites, denoted by A, A$^\prime$, B, and B$^\prime$, as shown in Fig.\,\ref{NoSOC1}(b). The primitive lattice vectors are given by $\textbf{a}_{1}=b \hat{x}$ and $\textbf{a}_{2}=a \hat{y}$,
where $b$ and $a$ are the lattice parameters in $\hat{x}$ ({\it zigzag}) and $\hat{y}$ ({\it armchair})  direction, respectively. Note that for a given phosphorus atom A (A$^\prime$) there are two nearest neighbor B(B$^\prime$) atoms in the same plane, and only one nearest neighbor A$^\prime$(A) atom out of the plane. Therefore the relative position vectors between the nearest neighbor atoms can be written as
\begin{equation}\label{distances}
\begin{aligned}
\bm{\delta}_{{\rm AB}_{\mathit{l}}} & =b_{1}\left(n_{lx},n_{y},n_{z}\right),\\
\bm{\delta}_{\rm A'B'_{\mathit{l}}} & = b_{1}\left( n_{lx},-n_{y},n_{z}\right),\\
\bm{\delta}_{\rm AA'} & = b_{2}(m_{x},-m_{y},-m_{z}),\\
\bm{\delta}_{\rm BB'} & = b_{2}(m_{x},m_{y},-m_{z})
\end{aligned}
\end{equation}
\noindent with the definitions,
\begin{gather}
n_{lx}=(-1)^{\mathit{l}}n_{x}(\theta,\mbox{\small $ \frac{\pi}{2}$}), \quad n_{y}=n_{y}(\theta,\mbox{\small $ \frac{\pi}{2}$}),  \quad n_{z}=n_{z}(\mbox{\small $ \frac{\pi}{2}$}), \nonumber \\ 
 m_{x}=n_{x}(0,\varphi), \quad  m_{y}=n_{y}(0,\varphi), \quad m_{z}=n_{z}(\varphi),
\end{gather}
\noindent where $b_{1}$ is the A(A$^\prime$)--B(B$^\prime$) atomic distance, while $b_{2}$ is the A(B)--A$^\prime$(B$^\prime$) atomic separation. Here $l=1,2$ and distinguishes the first two, in-plane, neighbors ${\rm B(B')}$ for the atom ${\rm A(A')}$. The cosine directors which in terms of the Euler angles of the line joining the ${\rm A(A')-B(B')}$ atomic sites with respect to the $\hat{y}$ and $\hat{z}$ axes are given by,  $n_{x}(\theta,\varphi) = \sin{\theta}\sin{\varphi}$, $n_{y}(\theta,\varphi) = \cos{\theta}\sin{\varphi}$, and $n_{z}(\varphi) = \cos{\varphi}$.
The values of all the parameters mentioned above have been taken by relaxed  DFT \cite{
Kurpas2016} calculations, and are reproduced in Table\,\ref{param1}. As the unit cell in reciprocal space for monolayer phosphorene is also rectangular (Fig.\,\ref{NoSOC1}(c)),  thus its the first Brillouin zone has a rectangular shape with its high symmetry points denoted by $\rm{S}$, $\rm{X}$, $\rm{Y}$, and $\Gamma$, which are located at the points $(\pi/b,\pi/a)$, $(\pi/b,0)$, $(0,\pi/a)$ and $(0,0)$, respectively.
\begin{table}[H]
\centering
\begin{tabular}{cc}
\hline \hline
$b_{1}$  & 2.222 [\r{A}]  \\
$b_{2}$  & 2.260  [\r{A}]  \\
$\theta$ & $47.92^{\circ}$\\
$\varphi$ & $21.29^{\circ}$\\
{\it a}  & 4.620  [\r{A}] \\
{\it b}   & 3.298  [\r{A}]  \\
\hline \hline
\end{tabular}
\caption{Relaxed geometrical parameters of monolayer black phosphorus.\cite{Kurpas2016}} 
\label{param1}
\end{table}

\begin{table}[!htb]
    \centering
     
    \begin{tabular}{rSS} 
\hline\hline

\multicolumn{1}{l}{\text{SK-integral}} & \text{Without SOC  (eV)} & \text{ With SOC (eV)} \\
\hline
$V_{pp\sigma}$ &  3.85 & 3.81     \\
$V_{pp\pi}$ & -0.99 & -0.95   \\
$V_{\rm{AB}}$ & -1.02 & -0.98  \\
$V_{\rm{AA}}$ & -0.39 & -0.38   \\
$V_{sp\sigma}$ & 2.16 & 2.15   \\
$V_{ss\sigma}$ & -1.99 & -1.99   \\
$\varepsilon_{s}$ & -12.03 & -11.87  \\
$\varepsilon_{p}$ & -3.42 & -3.33  \\
$\xi_p$ &   & 0.047     \\
$\Delta_{z}$ &   & 0.23  \\ 
\hline\hline
\end{tabular}
 
\caption{\label{paramSOC} Slater and Koster (SK) parameters obtained here from the DFT+W90 calculations without and with SOC. 
}
\end{table}

\section{{\it ab initio} computational method } \label{DFTsection}

First principles simulations were performed using Density Functional Theory (DFT) as implemented in the Vienna Ab-initio Simulation Package \cite{vasp} (VASP 5.4.4) using the relaxed structure as provided by Ref.[\onlinecite{Kurpas2016}]. The hybrid Becke three-parameter Lee-Yang-Parr (B3LYP) approximation \cite{Stephens1994} was used to account for exchange and correlation effects, which have been shown to accurately reproduce the bandgap energy in phosphorene \cite{Frank2019}. The lattice constants were selected based on the relaxed structural parameters provided in Table\,\ref{param1}. The interaction between ions and electrons was described using the projector-augmented wave (PAW) method \cite{PAW1994}. DFT calculations were carried out using an 11$\times$9$\times$1 $\Gamma$-centered $k$-mesh and a plane wave energy cutoff of 420 eV. Additionally, an augmentation charge cutoff energy of 343 eV was utilized in the calculations. 
SOC effects was included within the PAW formalism in the self consistent DFT method \cite{SOCVASP}.

To construct the effective Wannier Hamiltonian in the maximally localized Wannier basis, atomic orbitals (P:$s$;$p$) were projected onto using the WANNIER90 code \cite{Mostofi2014} as a post-processing step of the DFT calculations. The Wannier Hamiltonian was derived within a localized orbital basis set without the need for additional wannierization procedures. This Hamiltonian facilitates the extraction of tight-binding Slater-Koster parameters, simplifying the description of the electronic structure of phosphorene.

\section{Effective Hamiltonian in reciprocal space without spin-orbit coupling}\label{CH}

The explicit form of matrix elements of the Hamiltonian $H_{\phi}$ in Eq.\,(\ref{HphiR}) are given by,
\begin{equation}\label{recspace}
\begin{split}
H_{{\rm AA}'} &= e^{i{\bm k}\cdot {\bm \delta_{\rm{AA'}}}} \bra{\phi^{\rm A}}\hat{H}_0\, |{\phi^{\rm A'}}\rangle = V_{pp\sigma} e^{-ik_{y}h}, \\
H_{\rm{AB}} &= \sum_{l} e^{i{\bm k}\cdot {\bm \delta_{\rm{AB}_{\mathit{l}} } }}\bra{\phi^{\rm A}}\hat{H}_0\ket{\phi^{{\rm B}_{l}}}\\ &= 2V_{\rm{AB}}\, e^{iyk_{y}} \cos(\frac{bk_{x}}{2}),\\
H_{\rm{A'B'}} &= \sum_{l} e^{i{\bm k}\cdot {\bm \delta_{\rm{A'B'}_{\mathit{l}}}}} \langle{\phi^{\rm A'}}|\hat{H}_0 | {\phi^{{\rm B}'_{\mathit{l}}}} \rangle\\ &= 2V_{\rm{AB}}\,e^{-iyk_{y}} \cos(\frac{bk_{x}}{2}),\\
H_{\rm{BB'}} &= e^{i{\bm k}\cdot {\bm \delta_{\rm{BB'}}}} \bra{\phi^{\rm B}}\hat{H}_0 |{\phi^{{\rm B}'}} \rangle = V_{pp\sigma} e^{ik_{y}h}.
\end{split}
\end{equation}

Now, in order to find the continuum Hamiltonian corresponding to the model without spin orbit coupling, we find convenient to first rewrite Eq.\,(\ref{HphiR}), in the basis $\{\ket{\phi^{\rm{A}}}, |\phi^{\rm{B'}}\rangle, |\phi^{\rm{A'}}\rangle, \ket{\phi^{\rm{B}}}\}$  which yields to
\begin{equation}\label{HphiRrot}
 H_{\phi,\rm{T}}(\bm k) = \begin{pmatrix}
 0 & 0 & g(\bm k) V_{pp\sigma} & f(\bm k) V_{\rm AB} \\
  0 & 0 & f(\bm k) V_{\rm AB} & g(\bm k) V_{pp\sigma} \\
   g^{\ast}(\bm k) V_{pp\sigma} & f^{\ast}(\bm k) V_{\rm AB} & 0 & 0 \\
   f^{\ast}(\bm k) V_{\rm AB} & g^{\ast}(\bm k) V_{pp\sigma} & 0 & 0   
\end{pmatrix},
\end{equation}

\begin{widetext}
that by taking its square leads to a block diagonal matrix,  
\begin{equation}\label{HphiRrot2}
 H^{2}_{\phi,\rm{T}}(\mathbf{k}) = \begin{pmatrix}
 V^2_{pp\sigma} + 2 V^2_{\rm{AB}} g_2({\bm k}) & 4 V_{\rm{AB}} V_{pp\sigma} f_2({\bm k}) & 0 & 0 \\
  4 V_{\rm{AB}} V_{pp\sigma} f_2({\bm k}) & V^{2}_{pp\sigma} + 2 V^{2}_{\rm{AB}} g_2({\bm k}) & 0 & 0 \\
   0 & 0 & V^{2}_{pp\sigma} + 2 V^{2}_{\rm{AB}} g_2({\bm k}) & 4 V_{\rm{AB}} V_{pp\sigma} f_2({\bm k}) \\
   0 & 0 & 4 V_{\rm{AB}} V_{pp\sigma} f_2({\bm k}) & V^{2}_{pp\sigma} + 2 V^{2}_{\rm{AB}} g_2({\bm k}) 
\end{pmatrix},
\end{equation}
\end{widetext}
where we have defined the functions, 
\begin{equation}
\begin{split}
    f_2({\bm k}) &= \cos(\frac{b k_x}{2}) \cos(\frac{a k_y}{2}), \\
    g_2({\bm k}) &= 2\cos^2{\left(\frac{bk_x}{2}\right)}.
\end{split}
\end{equation}
The eigenvalues of the squared Hamiltonian of Eq.\,(\ref{HphiRrot2}) are given by
\begin{equation}\label{epsilon2}
\begin{split}
   \varepsilon^2_{\mu}(\bm k) &= 2 V^{2}_{\rm{AB}} + V^{2}_{pp\sigma} + 2 V^{2}_{\rm{AB}} \cos{(bk_x)} \\ &-4(-1)^\mu V_{\rm{AB}} V_{pp\sigma} \cos{\Big(\frac{b k_x}{2}\Big)} \cos(\frac{a k_y}{2}),
   \end{split}
\end{equation}
hence the dispersions of the four bands are ${\cal E}_{\pm\mu}(\bm k)=\pm \sqrt{|\varepsilon_{\mu}|} $, where $\mu=1,2$ denotes the index of the band, and the sign $\pm$ for the conduction/valence  character ($c/v$) of a given band. The eigenvectors of (\ref{HphiRrot2}) are
\begin{equation}\label{eigenfH22}
\begin{split}
\ket{\psi_{v_2}} &= \frac{1}{\sqrt{2}} \begin{pmatrix}
 0 \\
 0 \\
 -1 \\
 1 \\
\end{pmatrix}, \quad\quad 
\ket{\psi_{c_2}} = \frac{1}{\sqrt{2}} \begin{pmatrix}
 -1 \\
  1 \\
  0 \\
  0
\end{pmatrix}, \\
\ket{\psi_{v_1}} &= \frac{1}{\sqrt{2}} \begin{pmatrix}
 0 \\
 0 \\
 1 \\
 1 \\
\end{pmatrix}, \quad\quad 
\ket{\psi_{c_1}} = \frac{1}{\sqrt{2}} \begin{pmatrix}
 1 \\
  1 \\
  0 \\
  0
\end{pmatrix}.
\end{split}
\end{equation}
\newline
which are utilized to construct a unitary matrix,
\begin{equation}\label{Unit}
\widetilde{U} = \frac{1}{\sqrt{2}} \begin{pmatrix}
 0 & -1 & 0 & 1 \\
 0 & 1 & 0 & 1 \\
 -1 & 0 & 1 & 0 \\
 1 & 0 & 1 & 0    
\end{pmatrix}.
\end{equation}
\newline
that led to a block-diagonal Hamiltonian $\widetilde{H}_{\phi, \rm{T}}$ through the similar transformation,
\begin{equation}\label{HphiRT}
\begin{split}
&\widetilde{H}_{\phi, \rm{T}} = \widetilde{U}^{\dagger} H_{\phi,\rm{R}} \widetilde{U} \\ &= \begin{pmatrix}
 0 & {\cal V}_{2}^{\ast}(\bm{k})  & 0 & 0 \\
 {\cal V}_{2}(\bm{k}) & 0 & 0 & 0 \\
 0 & 0 & 0 & {\cal V}_{1}^{\ast}(\bm{k}) \\
 0 & 0 & {\cal V}_{1}(\bm{k}) & 0    
\end{pmatrix},
\end{split}
\end{equation}
where the function ${\cal V_{\eta}}(\bm{k})=V_{\rm{AB}}f(\bm{k})+(-1)^{\eta+1}V_{pp\sigma}g(\bm{k})$, and $f(\bm k)$ and $g(\bm k)$ were defined in Eq.\,(\ref{spechtralf}).  Subsequently, The matrix Hamiltonian in Eq.\,(\ref{HphiRT}) can be transformed into a new basis $U$ that couples the conduction and valence block states for each given band index $\eta=1,2$.  Such unitary transformation has the form 
\begin{equation}\label{Unit2}
U = \frac{1}{\sqrt{2}} \begin{pmatrix}
 1 & 1 & 0 & 0 \\
 -1 & 1 & 0 & 0 \\
 0 & 0 & 1 & 1 \\
 0 & 0 & -1 & 1    
\end{pmatrix}.
\end{equation}
Hence, the Hamiltonian transformed to new the basis $\{\ket{\phi_{v2}}, |\phi_{c2}\rangle, \ket{\phi_{v1}}, |\phi_{c1}\rangle\}$ reads,
\begin{equation}
\label{H4X4}
{\cal H}^{\prime}=U^{\dagger}\widetilde{H}_{\phi, \rm{T}} U= \begin{pmatrix} 
{\cal H}_{2} & 0 \\
0 & {\cal H}_{1} 
\end{pmatrix},
\end{equation}
where,
\begin{equation}
\label{H2X2}
\begin{split}
 {\cal H}_{\eta}(\bm{k}) 
&= (-1)^\eta\begin{pmatrix}
  Re[{\cal V}_{\eta}(\bm{k})] &   i Im[{\cal V}_{\eta}(\bm{k})] \\
  -i Im[{\cal V}_{\eta}(\bm{k})] &  -Re[{\cal V}_{\eta}(\bm{k})]
\end{pmatrix}.
\end{split}
\end{equation}

and the explicit form for such basis  is given by,
\begin{equation}\label{eigenfHvalcond}
\begin{split}
\ket{\phi_{v1}} &= \frac{1}{2}(|\phi^{\rm{A'}}\rangle+ |\phi^{\rm{B}}\rangle -  |\phi^{\rm{A}}\rangle - |\phi^{\rm{B'}}\rangle), \\
\ket{\phi_{c1}} &= \frac{1}{2} (|\phi^{\rm{A'}}\rangle+ |\phi^{\rm{B}}\rangle +  |\phi^{\rm{A}}\rangle + |\phi^{\rm{B'}}\rangle), \\
\ket{\phi_{v2}} &= \frac{1}{2} (|\phi^{\rm{B}}\rangle- |\phi^{\rm{A'}}\rangle -  |\phi^{\rm{B'}}\rangle + |\phi^{\rm{A}}\rangle), \\
\ket{\phi_{c2}} &= \frac{1}{2} (|\phi^{\rm{B}}\rangle- |\phi^{\rm{A'}}\rangle +  |\phi^{\rm{B'}}\rangle - |\phi^{\rm{A}}\rangle).
\end{split}
\end{equation}

\section{Effective Hamiltonian with spin-orbit interaction}\label{SOC}

In this appendix we outline the derivation of the effective Hamiltonian model including the spin-orbit coupling effects. With this aim we first expand the original basis $|\phi\rangle$  introduced in Eq.(\ref{basisphi}), to include spin and
the hybridization of the in-plane $p_x$ and $p_y$-like orbitals, as well as the $s$-like orbitals; here denoted compactly by    $|\Psi\rangle=\{|\phi\rangle,|\chi\rangle\}\otimes \{\uparrow,\downarrow\}$, with 
\begin{equation}
|\chi\rangle=\{|\Theta^{\rm{A}}\rangle, |\Theta^{\rm{A'}}\rangle, |\Theta^{\rm{B}}\rangle, |\Theta^{\rm{B'}}\rangle, |s^{\rm{A}}\rangle, |s^{\rm{A'}}\rangle, |s^{\rm{B}}\rangle,|s^{\rm{B'}}\rangle\},\nonumber    
\end{equation}

\noindent in which the $ |\Theta^{\rm A/A'}\rangle$ and $ |\Theta^{\rm B/B'}\rangle$ are linear combinations of $p_x$ and $p_y$-like orbitals as established in Eq.(\ref{basistheta}). Therefore, the full $(24\times24)$ Hamiltonian in the basis $|\Psi\rangle$ has the form
 
\begin{equation}\label{CompleteH}
    \mathcal{H}_{T}=
    \begin{pmatrix}
    \mathcal{H}_{\phi} & \mathcal{U} \\
   \mathcal{U}^{\dagger}  & \mathcal{H}_{\Theta,s} 
  \end{pmatrix},
\end{equation}
where $\mathcal{H}_{\phi}=H_{\phi} \otimes \{\uparrow,\downarrow\}$ is a $(8\times8)$ matrix, with $H_{\phi}$ provided  in Eq.\,(\ref{Hphi}). $\mathcal{H}_{\Theta,s} =H_{\Theta,s} \otimes \{\uparrow,\downarrow\}$ is a $(16\times16)$ Hermitian matrix, where $H_{\Theta,s}$ contains the onsite and hopping integrals between the $\ket{\Theta}$ and $\ket{s}$  orbital states, and finally, the $(8\times 16)$ complex matrix $\mathcal{U}$ and the $(16\times 16)$ $\mathcal{U}^{\dagger}$ contain the hopping integrals between $\ket{\phi} \otimes \{\uparrow,\downarrow\}$ and $\ket{\chi} \otimes \{\uparrow,\downarrow\}$ states.
Explicitly, the Hamiltonian $H_{\Theta,s}$ in the basis $\{\ket{\Theta^{\rm{A}}}, |\Theta^{\rm{A'}}\rangle, \ket{\Theta^{\rm{B}}}, |\Theta^{\rm{B'}}\rangle, \ket{s^{\rm{A}}}, |s^{\rm{A'}}\rangle, \ket{s^{\rm{B}}}, |s^{\rm{B'}}\rangle\}$ is found to be, 

\begin{widetext}
\begin{equation}\label{Htheta}
    H_{\Theta,s} = \begin{pmatrix}
    0 & V_{\rm{AA}} & -V_{pp\sigma} & 0 & 0 & -n_{y} m_y V_{sp\sigma} & -V_{sp\sigma} & 0 \\
   V_{\rm{AA}}  & 0 & 0 & -V_{pp\sigma} & -n_{y} m_y V_{sp\sigma} & 0 & 0 & -V_{sp\sigma} \\
   -V_{pp\sigma} & 0 & 0 & V_{\rm{AA}} & -V_{sp\sigma} & 0 & 0 & -n_{y} m_y V_{sp\sigma}\\
   0 & -V_{pp\sigma} & V_{\rm{AA}} & 0 & 0 & -V_{sp\sigma} & -n_{y} m_y V_{sp\sigma} & 0 \\
   0 & -n_{y} m_y V_{sp\sigma} & -V_{sp\sigma} & 0 & \varepsilon_{sp} & V_{ss\sigma} & V_{ss\sigma} & 0\\
   -n_{y} m_y V_{sp\sigma} & 0 & 0 & -V_{sp\sigma} & V_{ss\sigma} & \varepsilon_{sp} & 0 & V_{ss\sigma} \\
   -V_{sp\sigma} & 0 & 0 & -n_{y} m_y V_{sp\sigma} & V_{ss\sigma} & 0 & \varepsilon_{sp} & V_{ss\sigma} \\
   0 & -V_{sp\sigma} & -n_{y} m_y V_{sp\sigma} & 0 & 0 & V_{ss\sigma} & V_{ss\sigma} & \varepsilon_{sp} 
  \end{pmatrix}, 
\end{equation}
%
where the atomic on site energy difference between the $s$ and $p$ states is $\varepsilon_{sp} = \varepsilon_{s} - \varepsilon_{p}$, and 
we have found that the matrix elements  
$\langle \Theta^{\rm{A/A'}}|\hat{H}_{0}|\Theta^{\rm{B/B'}}\rangle = -V_{pp\sigma}$ and that $V_{\rm{AA'}} = V_{\rm{BB'}} $ with
\begin{equation}\label{VAA}
\begin{split}
V_{\rm{AA'}} &= \langle \Theta^{\rm{A/B}}|\hat{H}_{0}|\Theta^{\rm{A'/B'}}\rangle \\ &=n^2_{lx}V_{pp\pi} - n^2_y (m^2_y V_{pp\sigma} + (1 - m^2_y) V_{pp\pi}),
\end{split}
\end{equation}
the remainding SK parameters, $V_{ss\sigma}$ and $V_{sp\sigma}$ are defined in Table\,\ref{paramSOC}. 
The coupling matrix, $\mathcal{U}$, between the $\ket{\phi} \otimes \{\uparrow,\downarrow\}$ and $\ket{\chi} \otimes \{\uparrow,\downarrow\}$ are povided in Table\,\ref{matrixU} below,

\begin{table}[H]
\centering
\begin{tabular}{c|c c c c c c c c}
\vspace{0.1cm}
 & $\ket{\Theta^{\rm{A}},\uparrow\downarrow}$ & $|\Theta^{\rm{A'}},\uparrow\downarrow\rangle$ & $\ket{\Theta^{\rm{B}},\uparrow\downarrow}$ & $|\Theta^{\rm{B'}},\uparrow\downarrow\rangle$ & $\ket{s^{\rm{A}},\uparrow\downarrow}$ & $|s^{\rm{A'}},\uparrow\downarrow\rangle$ & $\ket{s^{\rm{B}},\uparrow\downarrow}$ & $|s^{\rm{B'}},\uparrow\downarrow\rangle$ \\
\hline
$\langle{\phi^{\rm{A}},\uparrow\downarrow}|$ & $i u^{-}_{1}\xi_p$ & $V^{*}_{pp\sigma} s_{\rm o}$  & $V^{*}_{pp\sigma} s_{\rm o}$ & 0 & $\Delta_{z} s_{\rm o}$ & $V_{sp\sigma} s_{\rm o}$ & $V^{*}_{sp\sigma} s_{\rm o}$ & 0 \\
   
$\langle{\phi^{\rm{A'}},\uparrow\downarrow}|$ & $-V^{*}_{pp\sigma} s_{\rm o}$ & $i u^{+}_{1}\xi_p$ & 0 & $-V^{*}_{pp\sigma} s_{\rm o}$ & $-V_{sp\sigma} s_{\rm o}$ & $\Delta_{z} s_{\rm o}$ & 0 & $-V^{*}_{sp\sigma} s_{\rm o}$ \\

$\langle{\phi^{\rm{B}},\uparrow\downarrow}|$ &   $V^{*}_{pp\sigma} s_{\rm o}$ & 0 & $i u^{+}_{2}\xi_p$ & $V^{*}_{pp\sigma} s_{\rm o}$ & $V^{*}_{sp\sigma} s_{\rm o}$ & 0 & $\Delta_{z} s_{\rm o}$ &  $V_{sp\sigma} s_{\rm o}$ \\
   
$\langle{\phi^{\rm{B'}},\uparrow\downarrow}|$ & 0 & $-V^{*}_{pp\sigma} s_{\rm o}$ & $-V^{*}_{pp\sigma} s_{\rm o}$ & $i u^{-}_{2}\xi_p$ & 0 & $-V^{*}_{sp\sigma} s_{\rm o}$ & $-V_{sp\sigma} s_{\rm o}$ & $\Delta_{z} s_{\rm o}$\\
\hline
\end{tabular}
\caption{Elements of the coupling matrix $\mathcal{U}$. The short notation $\ket{\Theta^{\rm{A}},\uparrow\downarrow}$  represent the set $\{\ket{\Theta^{\rm{A}}\uparrow },\ket{\Theta^{\rm{A}}\downarrow}\}$, and similarly for the other states. }
\label{matrixU}
\end{table}

    
\noindent where $V^{*}_{pp\sigma} = n_{y} m_y V_{pp\sigma}$, and $V^{*}_{sp\sigma} = n_{y} m_y V_{sp\sigma}$, and we have defined $u^{\pm}_{1} = -n_{lx} m_{y}s_z + (n_{lx} s_y \pm n_{y} s_x)m_z$ and $u^{\pm}_{2} = -n_{lx} m_{y} s_z - (n_{lx} s_y \mp n_{y} s_x)m_z$, being $\{s_x, s_y, s_z\}$ the spin Pauli matrices, and $s_{\rm o}$ is the $2\times 2$ identity matrix. Starting from Eq.(\ref{CompleteH}) an effective Hamiltonian in real space can be obtained  by using the band-folding method considering the $\ket{\chi} \otimes \{\uparrow,\downarrow\}$  states in perturbation theory,
\begin{equation}\label{Heffphitheta}
\begin{split}
    \mathcal{H}_{\rm eff} \approx \mathcal{H}_{\phi} - \mathcal{U} \mathcal{H}^{-1}_{\Theta,s} \mathcal{U}^{\dagger} = \begin{pmatrix}
     0 & V_{pp\sigma} s_{o} +i \Lambda_{R+} s_x & V_{\rm{AB}} s_{o} -i\Lambda_{R} s_{-} & -i\Lambda_{I}s_{z} \\
    V_{pp\sigma} s_{o} -i \Lambda_{R+} s_x  & 0 & i\Lambda_{I}s_z & V_{AB} s_{o} +i\Lambda_{R} s_{+} \\
    V_{\rm{AB}} s_{o} +i\Lambda_{R} s_{-} & -i\Lambda_{I}s_{z} &  0 & V_{pp\sigma} s_{o} -i\Lambda_{R-} s_x \\
   i\Lambda_{I}s_{z} & V_{\rm{AB}} s_{o} -i\Lambda_{R} s_{+} & V_{pp\sigma} s_{o} +i\Lambda_{R-} s_x & 0
  \end{pmatrix},
\end{split}
\end{equation}
%
with $s_{\pm}=n_{lx} s_y \pm n_{y} s_x$, and,
\begin{equation} \label{lambdas}
    \begin{split}
       \Lambda_I &= \frac{2 n_{lx} m_y n^2_y V_{pp\sigma} \xi_p}{V_{\rm{AA}} - V_{pp\sigma}},  \\
       \Lambda_{R\pm} &= \frac{2 m_z n_{y} V_{sp\sigma} V_{ss\sigma} \Delta_{st} \xi_p (1 \pm n_y m_y)} { (V_{\rm{AA}} - V_{pp\sigma}) \varepsilon^2_{sp}}\Big(1 + \frac{4 V^2_{ss\sigma}}{\varepsilon^2_{sp}}\Big), \\
        \Lambda_{R} &= \frac{2 m_z V_{sp\sigma} V_{ss\sigma}  \Delta_{st} \xi_p (1 + n_y m_y)}{ (V_{\rm{AA}} - V_{pp\sigma}) \varepsilon^2_{sp}}.
    \end{split}
\end{equation}

Note that the coupling parameter $ \Lambda_I$ arises due the intrinsic SOC through $\xi_p$, whereas the coefficients   
$\Lambda_{R}$ and $\Lambda_{R_\pm}$ depend on both, $\xi_p$ and $\Delta_z$, being the latter proportional to the electric field, which suggests its Rashba-type nature. 

\section{Effective Hamiltonian with spin-orbit interaction in reciprocal space}\label{SOCR}

Once the Fourier transform in $\bm k$-space is perfomed to the expression in Eq.\,(\ref{Heffphitheta}), the following effective Hamiltonian is obtained
\begin{equation}\label{Heffphithetak}
\begin{split}
    \mathcal{H}({\bm k}) = 
    &\begin{pmatrix}
     0 & (V_{pp\sigma} s_{o} +i \lambda_{R+} s_x) g({\bm k}) & V_{\rm{AB}} s_{o} f({\bm k}) - i\lambda_{R} S_{xy}({\bm k}) & \lambda_{I} s_{z} h^{\ast}({\bm k}) \\
    (V_{pp\sigma} s_{o} -i\lambda_{R+} s_x)g^*({\bm k})  & 0 & -\lambda_{I} s_{z} h({\bm k}) &  V_{\rm{AB}} s_{o} f^*({\bm k}) - i\lambda_{R} S_{xy}({\bm k}) \\
    V_{\rm{AB}} s_{o} f^*({\bm k}) + i\lambda_{R} S_{xy}^{*}({\bm k}) & -\lambda_{I} S_{z} h^{\ast}({\bm k}) &  0 & (V_{pp\sigma} s_{o} +i\lambda_{R-} s_x) g^*({\bm k}) \\
   \lambda_{I} s_{z} h({\bm k}) & V_{\rm{AB}} s_{o} f({\bm k}) + i\lambda_{R} S_{xy}^{*}({\bm k}) & (V_{pp\sigma} s_{o} -i\lambda_{R-} s_x) g({\bm k}) & 0
  \end{pmatrix},
\end{split}
\end{equation}

\noindent which is written in the basis $\{\ket{\phi^{\rm{A}},\uparrow\downarrow}, |\phi^{\rm{A'}},\uparrow\downarrow\rangle, \ket{\phi^{\rm{B}},\uparrow\downarrow}, |\phi^{\rm{B'}},\uparrow\downarrow\rangle\}$, with $S_{xy}({\bm k})=   s_x \cos\theta f({\bm k}) - i s_y \sin\theta f^{\prime}({\bm k})$, and

\begin{equation} \label{Lambdas}
    \begin{split}
       \lambda_I &= \frac{2 V_{pp\sigma} \xi_p \cos\theta \sin{\theta} \sin^2{\varphi}} {V_{\rm{AA}} - V_{pp\sigma}},  \\
       \lambda_{R\pm} &= \frac{2 V_{ss\sigma} V_{sp\sigma} \Delta_{st} \xi_p \cos{\theta} \cos{\varphi} (1 \pm \cos\theta\sin\varphi)}{(V_{\rm{AA}} - V_{pp\sigma}) \varepsilon^2_{sp}} \Big(1 + \frac{4 V^2_{ss\sigma}}{\varepsilon^2_{sp}} \Big),\\
        \lambda_{R} &= \frac{2 V_{sp\sigma} V_{ss\sigma} \Delta_{st} \xi_p \cos{\varphi} (1 + \cos\theta\sin\varphi) }{(V_{\rm{AA}} - V_{pp\sigma}) \varepsilon^2_{sp}},
    \end{split}
\end{equation}
\end{widetext}
where  $f(\bm k)$ and $g(\bm k)$ were defined in Eq.\,(\ref{spechtralf}), and we have introduced the new spectral functions, 
\begin{equation}\label{spechtralf2}
\begin{aligned}
{f^{\prime}}(\bm k) &= 2e^{i k_{y}y} 
\sin(\frac{bk_{x}}{2})\\
h(\bm k) &= 2\sin(\frac{bk_{x}}{2}) e^{ik_{y} a/2}.
\end{aligned}
\end{equation}

Observe that at $\Gamma$-symmetry point the spectral function $h(\bm k)$ vanishes, hence all the off-diagonal matrix elements of (\ref{Heffphithetak}) are identically zero. This leaves a Hamiltonian with a dominant Rashba-SOC and absent intrinsic SOC effects. 

\subsection{Low energy effective Hamiltonian with SOC at the $\Gamma$-point}\label{SOCRG}

We start by writing the Hamiltonian of Eq.\,(\ref{H4X4})  expanded into the spin basis  $\{ \ket{\phi_{v2}\uparrow},|{\phi}_{c2}\uparrow\rangle, \ket{\phi_{v2}\downarrow}, |{\phi}_{c2}\downarrow\rangle,\ket{\phi_{v1}\uparrow},|{\phi}_{c1}\uparrow\rangle, \ket{\phi_{v1}\downarrow}, |{\phi}_{c1}\downarrow\rangle\}$, 
\begin{equation}
{\cal H}_{0}= \begin{pmatrix} 
 H_{2} & 0_{4\times 4}\\
0_{4\times 4} & H_{1} 
\end{pmatrix}, \quad   H_{\mu}= \begin{pmatrix} 
{\cal H}_{\mu} & J \\
J^{*} & {\cal H}_{\mu} 
\end{pmatrix}, \quad \mu=1,2
\end{equation}
\begin{equation}
 J = \begin{pmatrix} 
{\cal I}_1(\bm{k}) & {\cal I}_2(\bm{k}) \\
{\cal I}^{*}_2(\bm{k}) & -{\cal I}_1(\bm{k}) 
\end{pmatrix},
\end{equation}

\noindent having defined
\begin{equation}
\begin{split}
    {\cal I}_1(\bm{k}) = & -\frac{1}{2} (\lambda_{R+} -\lambda_{R-}) \sin(h k_y)\\
    &- 2i \sin\theta \lambda_{R} \sin(\frac{b k_x}{2}) \cos(y k_y),
\end{split}
\end{equation}
    \begin{equation}
\begin{split}
    {\cal I}_2(\bm{k}) = & -\frac{1}{2} i (\lambda_{R+} -\lambda_{R-}) \cos(h k_y)\\
    &+ 2 \sin\theta \lambda_{R} \sin(\frac{b k_x}{2}) \sin(y k_y).
\end{split}
\end{equation}

Which, at linear order in $k_x$ and $k_y$, simplifies to
\begin{equation} \label{lambdasxy}
\begin{split}
{\cal I}_1(\bm{k}) &\approx -i (\lambda_{R} \sin\theta) b k_x - \frac{\lambda_{R+}-\lambda_{R-}}{2} h k_y \\
{\cal I}_2(\bm{k}) &\approx - i \frac{\lambda_{R+}-\lambda_{R-}}{2}.
\end{split}
\end{equation}


\begin{widetext}
\noindent hence for the lowest energy bands ($\eta =1$) the corresponding effective $\bm{k}\cdot \bm{p}$-type Hamiltonian ($H_1(\bm k)\rightarrow H(\bm k))$ that includes the Rashba SOC around the $\Gamma$ point reduces to 

\begin{equation}\label{HkGapp}
\begin{split}
    &H({\bm k}) \simeq\begin{pmatrix}
     (\alpha k_x^2 + \beta k_y^2 -\Delta) s_o + \lambda_{Rx} k_x s_y - \lambda_{Ry} k_y s_x & i v_y k_y s_o - i \Delta_{R_o} s_x  \\
    -i v_y k_y s_o + i \Delta_{R_o} s_x &  (-\alpha k_x^2 - \beta k_y^2 +\Delta) s_o - \lambda_{Rx} k_x s_y + \lambda_{Ry} k_y s_x
  \end{pmatrix},
\end{split}
\end{equation}

\noindent where we have defined the constants, 
\begin{equation}
\begin{split}
\lambda_{Rx} & = \lambda_{R} b \sin\theta  \simeq 4 b_1 \zeta_0  {(\sin{\theta})}^2,\\
\lambda_{Ry} & = \frac{h(\lambda_{R+}-\lambda_{R-})}{2}  \simeq  2 b_2 \zeta_0 {(\cos{\theta}\sin{\varphi})}^2 , \\
\Delta_{R_o} &= \frac{\lambda_{R+}-\lambda_{R-}}{2}=\frac{\lambda_{Ry}}{b_2 \sin{\varphi} }.
\end{split}
\end{equation}

\noindent with
\begin{equation}\label{zeta0A}
\zeta_0 = \frac{ \Delta_{z} \xi_p V_{ss\sigma}V_{sp\sigma} \cos{\varphi} }{(V_{\rm AA} - V_{pp\sigma})(\varepsilon_s -\varepsilon_p)^2}. 
\end{equation} 
\noindent Expression (\ref{HkGapp}) correspond to the Hamiltonian provided compactly in  Eq.\,\eqref{HGSOC}.  The eigenenergies of (\ref{HkGapp}) yields,
\begin{equation}
\label{EigenHRSO}
{\cal E}_{\sigma}^{\mu}(\bm k)=\mu \sqrt{v^2k_y^2+\Omega_k^2+\kappa^2 + \Delta_{R_0}^2+2\sigma \sqrt{\lambda_{R_x}^2 k_x^2(\Omega_k^2+\Delta_{R_0}^2) +\lambda_{R_y}^2 k_y^2\left(\Omega_k +\frac{v \Delta_{R_0}}{\lambda_{R_y}}\right)^2} }, 
\end{equation}
\begin{equation}
\text{with} \quad \Omega_k=\alpha k_x^2+\beta k_y^2 -\Delta\, , \quad \text{and}\quad \kappa^2=\lambda_{R_x}^2 k_x^2+\lambda_{R_y}^2 k_y^2, 
\end{equation} 

\noindent where $\mu=\pm$ labels the conduction/valence band and $\sigma=\pm$ its spin state. In the limit $\Delta_{R_o}\rightarrow 0$, the bandstructure (\ref{EigenHRSO}) reduces to Eq.(\ref{eigenSOCsimp}). 
\end{widetext}

\subsection{Low energy effective Hamiltonian with SOC at the ${\rm S}$-point}\label{SOCRS}

At the vecinity of the $\rm S$-point we have ${\bm k}= {\bm k}_s +{\bm q}$, being ${\bm q} =(\pi/b,\pi/a)$. At this point, is the intrinsic SOC that dominates over the Rashba-SOC. Hence we can safely preserve only the intrinsic type of SOC matrix elements in (\ref{Heffphithetak}).  That is, those matrix elements which are proportional to $\lambda_I$. Bearing this in mind, afterwards we can conveniently write the resulting Hamiltonian in proper basis that renders the Hamiltonian block-diagonal in such basis, leading to the Hamiltonian, 

\begin{equation}
\label{HISOfull}
{\cal H}(\bm k) = {\cal H}^{\prime}(\bm k)s_o + {\cal H}_{I}(\bm k),
\end{equation}

\noindent written in the basis $\{(\ket{\phi_{v2}}, |\phi_{c2}\rangle, \ket{\phi_{v1}}, |\phi_{c1}\rangle)\}\otimes |\uparrow , \downarrow\rangle$,  where ${\cal H}^{\prime}(\bm k)$ is given in (\ref{H4X4}), and the intrinsic spin-orbit term reads

\begin{equation}
\label{HeffphithetakS}
\begin{split}  
{\cal H}_I(\bm k)& = \begin{pmatrix}
H_{vc}^{\uparrow}  & 0   \\
 0   & H_{vc}^{\downarrow}
\end{pmatrix},
\end{split} 
\end{equation}
\begin{widetext}
with
\begin{equation}
H_{vc}^{\uparrow}=
\left(
\begin{array}{cccc}
 0 & -\lambda_{I}Re[h(\bm{k})] & -i\lambda_{I} Im[h(\bm{k})] & 0 \\
 \lambda_{I}Re[h(\bm{k})]& 0 & 0 & -i\lambda_{I} Im[h(\bm{k})]  \\
i\lambda_{I} Im[h(\bm{k})]  & 0 & 0 & \lambda_{I}Re[h(\bm{k})] \\
 0 & i\lambda_{I} Im[h(\bm{k})]  & \lambda_{I}Re[h(\bm{k})] & 0 \\
\end{array}
\right), \quad \text{and}\quad H_{vc}^{\downarrow}=-H_{vc}^{\uparrow}.
\end{equation}

\noindent Next, by avaluating (\ref{HISOfull}) in ${\bm k}= {\bm k}_s +{\bm q}$, and expanding the Hamiltonian around ${\bm k}_s$  whiles considering only the dominant terms  in ${\cal H}_I(\bm k)$, the eigenvalues of (\ref{HISOfull}) leads to the bandstructure at the $\rm S$-symmetry point reported in (\ref{EivaloS}), in which we have replaced there ${\bm k}_s\rightarrow {\bm k}$ to easy the notation. 

\end{widetext}
 

\end{document}